\documentclass[twocolumn,superscriptaddress,showpacs,prd,aps,amsmath,amssymb,nofootinbib]{revtex4-1}
\usepackage{natbib}
\usepackage{graphicx,color}
\usepackage{amsmath,amssymb,bigints}
\usepackage{verbatim}
\usepackage{float}
\usepackage{wasysym}
\usepackage{amssymb,graphicx}
\usepackage{epsfig}
\usepackage{psfrag}
\usepackage{dsfont}
\usepackage{amsfonts}
\usepackage{mathrsfs}
\usepackage{multirow}
\usepackage{times}
\usepackage{bm}
\usepackage{hyperref}
\usepackage{adjustbox}
\hypersetup{
  colorlinks=true,        
  linkcolor=blue,         
  citecolor=cyan,         
}
\begin{document}

\title{Black hole-neutron star coalescence: Effects of the neutron star spin on
 jet launching and dynamical ejecta mass}
\author{Milton Ruiz}
\affiliation{Department of Physics, University of Illinois at
  Urbana-Champaign, Urbana, IL 61801}
\author{Vasileios Paschalidis} \affiliation{Departments of Astronomy
  and Physics, University of Arizona, Tucson, AZ 85719}
\author{Antonios Tsokaros}
\affiliation{Department of Physics, University of Illinois at
    Urbana-Champaign, Urbana, IL 61801}
\author{Stuart L. Shapiro}
\affiliation{Department of Physics, University of Illinois at
  Urbana-Champaign, Urbana, IL 61801}
\affiliation{Department of Astronomy \& NCSA, University of
  Illinois at Urbana-Champaign, Urbana, IL 61801}


\begin{abstract}
  Black hole-neutron star (BHNS) mergers are thought to be sources of
  gravitational waves (GWs) with coincident electromagnetic (EM)
  counterparts. To further probe whether these systems are viable
  progenitors of short gamma--ray bursts (sGRBs) and kilonovae, and
  how one may use (the lack of) EM counterparts associated with
  LIGO/Virgo candidate BHNS GW events to sharpen parameter estimation,
  we study the impact of neutron star spin in BHNS mergers.
  Using dynamical spacetime magnetohydrodynamic simulations of BHNSs
  initially on a quasicircular orbit, we survey configurations
  that differ in the BH spin ($a_{\rm BH}/M_{\rm BH}=0$ and $0.75$),
  the NS spin ($a_{\rm NS}/M_{\rm NS}=-0.17,\,0,\,0.23$ and $0.33$),
  and the binary mass ratio ($q\equiv M_{\rm BH}:M_{\rm NS}=3:1$ and
  $5:1$). The general trend we find is that increasing the NS prograde
  spin increases both the rest mass of the accretion disk onto the
  remnant black hole, and the rest mass of dynamically ejected
  matter. By a time~$\Delta t\sim 3500-5500M\sim
  88-138(M_{\rm NS}/1.4M_\odot)\,\rm ms$ after the peak gravitational
  wave amplitude, a magnetically--driven jet is launched only for
  $q=3:1$ regardless of the initial NS spin. The lifetime of the jets
  [$\Delta t\sim 0.5-0.8(M_{\rm NS}/1.4 M_\odot)\,\rm s$] and their
  outgoing Poynting luminosity [$L_{\rm Poyn}\sim 10^{51.5\pm
      0.5}\,\rm erg/s$] are consistent with typical sGRBs luminosities
  and expectations from the Blandford--Znajek mechanism. By the time
  we terminate our simulations, we do not observe either an outflow or
  a large-scale magnetic field collimation for the other systems we
  considered. The mass range of dynamically ejected matter is
  $10^{-4.5}-10^{-2}~(M_{\rm NS}/1.4M_\odot)M_\odot$, which can
  power kilonovae with peak bolometric luminosities~$L_{\rm knova}\sim
  10^{40}-10^{41.4}$~erg/s with rise times $\lesssim 6.5\,\rm
  h$ and potentially detectable by the LSST.
\end{abstract}

\maketitle

\section{Introduction}
We are in a golden era of gravitational wave (GW) physics where the
sensitivity of ground-based laser interferometers is rapidly
increasing. During the first observing run
$O1$~\cite{LIGOScientific:2018mvr}, only $3$ GW events from binary
black hole (BBH) mergers were detected. The $O2$ run observed the
first GWs (GW170817) from the inspiral of a stellar compact
binary~\cite{TheLIGOScientific:2017qsa}, in which at least one of the
companions was a neutron star. The progenitor of this event has been
officially classified as a merging binary neutron star (BNS) system,
although the possibility of a merging black hole-neutron star (BHNS)
progenitor cannot be excluded~(see e.g.~\cite{Yang:2017gfb,
  Hinderer:2018pei}). In addition, $7$ new BBHs were
detected~\cite{LIGOScientific:2018mvr}.  Finally, during $O3$, whose
sensitivity was increased by $\sim 50\%$ compared to $O1$, at least
$53$ GW event candidates have been
reported~\cite{alertsGWweb}.\footnote{Recently $26$ of these events
  have been officially confirmed, along with 13 new GW events reported
  for the first time in ~\cite{Abbott:2020niy}.} These GW events can
be classified as follows: a) $37$~ BBHs candidates; b) $7$ BNS
candidates. It should be noted that the progenitor of GW190425 is a
BNS system with a total mass of~$3.4^{+0.3}_{-0.1}M_\odot$, which is
significantly different from the known population of Galactic BNS
systems~(see~e.g.~\cite{Tauris:2017omb}); c) 4 events in the so-called
mass gap (compact objects with masses of $3-5\ M_\odot$--see~e.g.
\cite{Mandel:2015spa,Littenberg:2015tpa,Tsokaros:2019lnx}); and d)
5~BHNS candidates, of which only one event has been confirmed
(GW190814) and whose inferred individual masses are
$23^{+1}_{-0.9}M_\odot$ and
$2.59^{+0.08}_{-0.08}M_\odot$~\cite{Abbott:2020khf}.  It is worth
noting that, although this event is listed
in~\cite{alertsGWweb,Abbott:2020niy}~as a BHNS candidate with $>99\%$
probability, due to the lack of any EM counterpart or tidal signature,
the nature of the lighter companion is uncertain (see~e.g.
\cite{Tsokaros:2020hli,Most:2020bba,Godzieba:2020tjn,Tews:2020ylw}).
If indeed the lighter binary companion is a NS then this would be the
heaviest NS yet observed~\cite{Cromartie:2019kug}. If, on the other
hand, the binary companion is a BH then it would be the lightest BH
observed to date. Notice the mass of the lightest stellar-mass BH
candidate observed in X-rays is~$3.8^{+0.5}_{-0.5}
M_\odot$~\cite{Orosz2004ApJ}.

Intense EM counterpart-observing campaigns preceding/following GW
detections have led to the following EM detections: a) a weak
transient EM signal~\cite{connaughton, Connaughton:2018bab} (event
GW150914-GBM, observed only by Fermi) that lasted $1$ s and appeared
$0.4$ s after the detection of GW150914, the first event consistent
with the inspiral and merger of a BBH; b) a MeV-scale EM signal
lasting for $32$ ms and occurring $0.46\,\rm s$ before GW170104 (also
consistent with a BBH~\cite{2017PhRvL.118v1101A}), as reported by the
AGILE mission~\cite{Verrecchia:2017lya}); c) EM counterparts across
the spectrum reported by several observatories (see
e.g.~\cite{GBM:2017lvd,Monitor:2017mdv,Abbott:2017wuw,Chornock:2017sdf,Cowperthwaite:2017dyu,
  Kasen:2017sxr,Nicholl:2017ahq}) following the detection of GW170817.
Its association with the transient GRB170817A~\cite{Monitor:2017mdv}
and the kilonova AT 2017gfo/DLT17ck~\cite{Valenti:2017ngx} provides
the best direct observational evidence so far that at least some sGRBs
are indeed powered by BNS mergers, or by the merger of a stellar
compact binary where at least one of the companions is a NS (or a
hybrid star~\cite{Paschalidis:2017qmb,Essick:2019ldf}). The BNS-sGRB
connection was anticipated in~\cite{Pac86ApJ,EiLiPiSc,NaPaPi}, and
numerically demonstrated by self-consistent simulations in full
general relativistic magnetohydrodynamics (GRMHD) of merging
BNSs~\cite{Ruiz:2016rai,Ruiz:2017inq,Ruiz:2020via} and
BHNSs~\cite{prs15,Ruiz:2018wah}. GW170817 and its EM counterpart
signals have been used to impose some constraints on the physical
properties of a NS~(see~e.g.~\cite{Margalit:2017dij,
  Shibata:2017xdx,Ruiz:2017due,Rezzolla:2017aly,Most:2018hfd,
  TheLIGOScientific:2017qsa,
  Abbott:2018exr,Radice:2017lry,Bauswein:2017vtn,Raithel:2018ncd,Raithel:2019ejc}),
such as the maximum mass of a spherical NS, as well as its tidal
deformability, equation of state, and radius
(see~\cite{Raithel:2019uzi,Baiotti:2019sew,Horowitz:2019piw,GuerraChaves:2019foa,Radice:2020ddv,Chatziioannou:2020pqz}
for reviews); d) an EM flare observed by the Zwicky Transient
Facility, consistent with an ejected BBH merger remnant in the
accretion disk of an active galactic nucleus that may be associated
with GW190521~\cite{PhysRevLett.124.251102}; and e) a weak EM
transient (GRB190425 event) $\sim 0.5$ s after GW190814 as reported
in~\cite{Pozanenko:2019lwh}.

Unlike the EM counterparts associated with GW170817, the other candidate
EM counterparts were not confirmed by other observatories/satellites
operating at the same time. The absence of observable EM counterparts
from candidate BHNS mergers may question their role as progenitors of
the central engines that power sGRBs.  Yet, GRMHD simulations
in~\cite{prs15,Ruiz:2018wah} showed that BHNS remnants of~$q=3:1$
mergers can potentially launch magnetically-driven jets.  Now early
population synthesis studies found that the distribution of mass
ratios $q$ in BHNSs depends on the metallicity, and peaks at
$q=7:1$~\cite{Belczynski:2007xg,Belczynski10}, but more recent work
finds that it is generally less than $10:1$, and peaks at $q\approx
5:1$~\cite{Giacobbo:2018etu,Abbott:2020khf}.  As pointed out
in~\cite{Foucart:2012nc,Foucart:2018rjc}, the larger the mass ratio,
the higher the BH spin required for the NS companion to be tidally
disrupted before reaching the innermost stable circular orbit
(ISCO). So far in BBHs reported by the LIGO/Virgo scientific
collaboration, BHs have high mass and/or low spins (see e.g.~Table VI
in~\cite{Abbott:2020niy}).  If this trend continues for LIGO/Virgo
BHNSs, then it is expected that LIGO/Virgo BHNS remnants would
have negligible accretion disks and ejecta~\cite{Foucart:2020ats},
which might disfavor their role as progenitors of sGRBs and
kilonovae. However, the NS spin could have a strong impact on the
tidal disruption and dynamical ejection of matter, affecting both sGRB
and potential kilonovae signatures. It should be noted that the spins
of the binary companions are only weakly constrained by current GW
observations.

In this paper, we survey fully relativistic BHNS configurations on a quasicircular orbit
undergoing merger in which the BH and/or the NS companions are spinning. 
We address two questions:
a) {\it Can a moderate high-mass ratio BHNS binary be the progenitor of an engine that powers
  sGRBs?};
b) {\it Can the spin of a NS companion change the fraction of the dynamical ejection of matter
  that may drive potentially detectable kilonovae signatures?}

We consider BHNS configurations with mass ratios $q=3:1$ and
$q=5:1$. In the first case the BH spin is $a_{\rm BH}/M_{\rm
  BH}=0.75$, while in the latter one the BH is nonspinning.  The NS
spin has a spin $a_{\rm NS}/ M_{\rm NS}=-0.17,\, 0,\, 0.23$ or $0.33$.
In all cases, the star is threaded by a dynamically weak poloidal
magnetic field that extends from the stellar interior into the
exterior (as in a pulsar), and whose dipole magnetic moment is aligned
with the orbital angular momentum of the binary. For purposes of
comparison with our earlier studies~\cite{prs15,Ruiz:2018wah}, the NS
is modeled by a polytropic equation of state (EOS) with $\Gamma=2$.

We find that the late inspiral and merger phases of the above BHNS
binaries are roughly the same as
in~\cite{Etienne:2007jg,Etienne:2011ea}, where the magnetic field is
confined to the interior of the star. The fraction of the total
rest-mass NS outside the horizon varies from $\lesssim 1\%$ to $\sim
15\%$ depending strongly on the binary mass ratio.  The general trend
is that {\it increasing the NS prograde spin increases both the mass
  of the accretion disk remnant and the unbound material (ejecta). In
  addition, NS spin leads to GW dephasing, with higher prograde spin
  increasing the number of GW cycles}.

Consistent with our previous results in~\cite{Ruiz:2018wah}, we find
that by~$\Delta t\sim 3500-5500M\approx 88-138(M_{\rm
  NS}/1.4M_\odot)\,\rm ms$ following the GW peak emission a
magnetically--driven jet emerges from the BH + disk remnant of BHNSs
with mass ratio $q=3:1$ regardless of the initial NS spin. However,
the jet launching time depends strongly on the latter. As the initial NS
prograde spin increases, the effective ISCO decreases and the
separation at which the star is tidally disrupted increases. These two
effects induce long tidal tails of matter that result in more
baryon-loaded environments. Thus, stronger magnetic fields are
required to overcome the baryon ram-pressure, delaying the launch of
the jet while the fields amplify. Notice that jet launching may
not be possible for all EOSs if the matter fall-back timescale is
longer than the disk accretion timescale~\cite{Paschalidis:2016agf}.
The lifetime of the jet [$\Delta t\sim 0.5-0.8 (M_{\rm
    NS}/1.4M_\odot)\,\rm s$] and outgoing Poynting luminosity [$L_{\rm
    Poyn}\sim 10^{51.5 \pm 0.5}\,\rm erg/s$] are consistent with
typical sGRB (see e.g. \cite{Bhat:2016odd,
  Lien:2016zny,Svinkin:2016fho,Ajello:2019zki}), and with the
Blandford--Znajek (BZ)~\cite{BZeffect}
luminosities~\cite{MembraneParadigm}. These results are also
consistent with a simple, ``universal'' model for BH + disk remnants
proposed in~\cite{Shapiro:2017cny}.

The characteristic temperature of the disk remnant is $T\sim
10^{11}\,\rm K$ (or~$\sim 8.6$~MeV) and hence it may also emit a copious
amount of neutrinos with peak-luminosity of $10^{53}\,\rm
erg/s$~\cite{Just:2015dba,Kyutoku:2017voj}. However, as the lifetime
of this process might be too small to explain typical
sGRBs~\cite{Kyutoku:2017voj}, it has been suggested that BH + disk
remnants powering sGRBs may be dominated initially by thermal pair
production followed by the BZ process~\cite{Dirirsa:2017pgm}.

Finally, we find the dynamical ejection of matter is strongly affected
by the initial NS spin. It ranges between $10^{-4.5}$ and
$10^{-2}(M_{\rm NS}/1.4M_\odot)M_\odot$, and may induce kilonovae
signatures with peak bolometric luminosities~of~$L_{\rm knova}\sim
10^{40} -10^{41.4}$~erg/s and rise times $\lesssim 6.5\,\rm h$,
potentially detectable by the Large Synoptic Survey Telescope (LSST)
survey~\cite{Mandelbaum:2018ouv} out to $O(200)$ Mpc. Similar
conclusions were reached in eccentric BHNS mergers with spinning NSs
in~\cite{East:2015yea}.  These preliminary results suggest that
moderate high mass ratio BHNSs that undergo merger, where the NS
companion has a non-negligible spin, may give rise to detectable
kilonovae signatures even if magnetically-driven jets are absent.

The remaining sections of the paper are organized as follows: A short
summary of our numerical methods and their implementation, along with
our initial data and the grid structure used to solve the GRMHD
equations, are presented in Sec.~\ref{sec:Methods}.  We present our
results in Sec.~\ref{sec:results} and conclusions in
Sec.~\ref{sec:conclusion}.  Geometrized units ($G=c=1$) are adopted
throughout the paper except where stated explicitly.
%
%
\begin{center}
  \begin{table}[th]
    \caption{Initial properties of the evolved BHNS configurations. We
      list the mass ratio $q\equiv M_{\rm BH}:M_{\rm NS}$, where
      $M_{\rm BH}$ is the BH mass at infinite separation and $M_{\rm
        NS}$ the NS rest-mass~(see~\cite{TBFS06} for details), the BH
      spin~$a_{\rm BH}/M_{\rm BH}$, the NS spin~$a_{\rm NS}/M_{\rm
        NS}$, which is either aligned or antialigned (indicated with a
      $-$ sign) with respect to the total angular momentum of the
      system. $T/|W|$ and $P$ are the kinetic-to-binding-energy ratio
      and the rotation period in units of $(M_{\rm
        NS}/1.4M_\odot)\,\rm ms$ of the NS, respectively. The
      dimensionless ADM mass $\bar{M}\equiv\kappa^{-1/2}\,M$ (here $\kappa$
      is the polytropic gas constant), the ADM angular momentum $J$ of
      the system, and the orbital angular velocity $\Omega_0$. The
      label for each configuration includes successively: a mass ratio
      tag ($q=3$ or $q=5$), and a tag identifying the spin direction
      (m = antialigned or p = aligned) and its magnitude.  In all
      configurations, the NS companions have a nondimensional
      rest-mass~${\bar{M}_{\rm NS}}=0.15$, and the initial
      $M\,\Omega_0$ corresponds to an orbital separation of about
      $D_0\simeq 8.7M\sim 66(M_{\rm NS} /1.4M_\odot)\,\rm km$.
      \label{table:BHNS_ID}}
    \begin{adjustbox}{width=\linewidth,center}
    \begin{tabular}{ccccccccc}
      \hline\hline
          Model            & $q$   &  $a_{\rm BH}/M_{\rm BH}$ &  $a_{\rm NS}/M_{\rm NS}$ & $T/|W|$& $P$ (ms)$^\ast$ &  $\bar{M}^\dag$ & $J/M^2$ &  $M\,\Omega_0$\\  
          \hline
           q3NS0.0$^\ddag$ &  3:1  & 0.75 & 0.0    &0.0& 0.0    &   0.55  &1.09& 0.0328\\
           q3NSm0.17       &  3:1  & 0.75 &$-0.17$   &0.009& 3.2   &   0.55  &1.09& 0.0328\\
           q3NSp0.23       &  3:1  & 0.75 & 0.23   &0.016 & 2.4   &   0.55  &1.09& 0.0328\\
           \hline\hline
           q5NS0.00$^\ddag$ &  5:1  & 0.0  & 0.0    &0.0  & 0.0    &   0.83 &0.52& 0.0333\\
           q5NSm0.17       &  5:1  & 0.0  &$-0.17$   &0.009& 3.2  &   0.83 &0.52& 0.0333\\
           q5NSp0.23       &  5:1  & 0.0  & 0.23   &0.016& 2.4   &   0.83 &0.52& 0.0333\\
           q5NSp0.33       &  5:1  & 0.0  & 0.33   &0.15 & 1.9   &   0.83 &0.52& 0.0333\\
          \hline\hline
    \end{tabular}
        \end{adjustbox}
    \begin{flushleft}
      $^{\ast}${Normalized to $(M_{\rm NS}/1.4M_\odot)$. Note that the fastest known pulsar has a period of $1.40\,\rm ms$~\cite{Hessels:2006ze}.}\\
      $^{\dag}M\simeq5.1M_\odot(M_{\rm NS}/1.4M_\odot)$.\\
      $^{\ddag}$ Cases treated previously in~\cite{Ruiz:2018wah}.
  \end{flushleft}
  \end{table}
\end{center}

%
\section{Numerical schemes}
\label{sec:Methods}
The numerical methods used to evolve the BHNS binaries are the same as in~\cite{Ruiz:2018wah}.
Therefore, in this section we briefly introduce our notation and summarize our numerical
schemes, along with the initial data. We refer the reader to~\cite{Ruiz:2018wah} for further
details. 
%
\paragraph*{\bf Formulation and numerical scheme:}
We solve Einstein's equations for the gravitational field coupled to
the MHD equations for the matter and magnetic field using the
adaptive--mesh--refinement (AMR) {\tt Illinois GRMHD} code embedded in
the {\tt Cactus/Carpet} infrastructure~\cite{Cactus,Carpet}. This code
uses the Baumgarte--Shapiro--Shibata--Nakamura (BSSN)
formulation~\cite{Shibata95, BS} to evolve the metric, and employs
moving puncture gauge conditions cast in first order
form~\cite{Hinder:2013oqa}.  Additionally, the code solves the MHD
equations in a conservative formulation [see Eqs.(27)-(29)
  in~\cite{Etienne:2010ui}] using high-resolution shock-capturing
methods~\cite{Duez:2004nf}.  We set the damping parameter $\eta$
appearing in the shift condition to $\eta=3.3/M$ for BHNSs with mass
ratio $q=3:1$, and to $\eta=1.2/M$ for those with mass ratio $q=5:1$
(see~Table~\ref{table:BHNS_ID}). Here $M$ is the Arnowitt-Deser-Misner
(ADM) mass of the
system. Following~\cite{UIUCEMGAUGEPAPER,Farris:2012ux}, we use the
generalized Lorenz gauge to avoid the spurious magnetic fields between
AMR levels due to numerical interpolations. We set the damping
parameter~$\xi\sim5.5/M$ for configurations with mass ratio $q=3:1$,
and $\xi=6.4/M$ for those with $q=5:1$. Finally, we adopt a
$\Gamma$--law EOS $P=(\Gamma-1) \rho_0\,\epsilon$ with $\Gamma=2$,
which allows shock heating during the evolution. Here $P$ and $\rho_0$
are the pressure and the rest-mass density, respectively. As in
standard hydrodynamic and MHD simulations, we integrate the ideal
GRMHD equations everywhere, imposing a tenuous constant--density
atmosphere $\rho_{0,\,{\rm atm}}=10^{-10}\,\rho_0^{\rm max}(0)$, where
$\rho_{0}^{\rm max}(0)$ is the initial maximum value of the rest-mass
density of the NS.
\begin{figure}[th]
  \centering
  \includegraphics[width=0.50\textwidth]{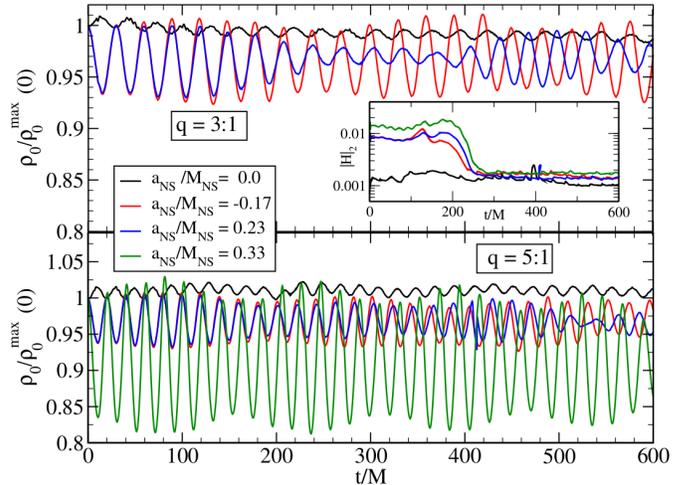}
  \caption{Maximum value of the rest-mass density $\rho_0(t)$ normalized to its initial maximum value~$\rho_0(0)$
    during the early inspiral for the nonmagnetized cases (see~Table \ref{table:BHNS_ID}). The inset shows the
    normalized $L_2$ norm of the Hamiltonian constraint~(see Eqs.~(40) and (41)~in~\cite{Etienne:2007jg}) for the
    $q=5:1$~cases. Due to the constraint damping used in our evolutions~(see Eq.~19 in~\cite{DMSB}),
    after about $t\sim 220M$ (or around one orbit), the Hamiltonian constraint falls
    roughly to the same low value for all cases in Table~\ref{table:BHNS_ID} regardless of the NS spin.
    \label{fig:rho_spin}}
\end{figure}

%
\begin{figure*}
  \centering
  \includegraphics[width=0.33\textwidth]{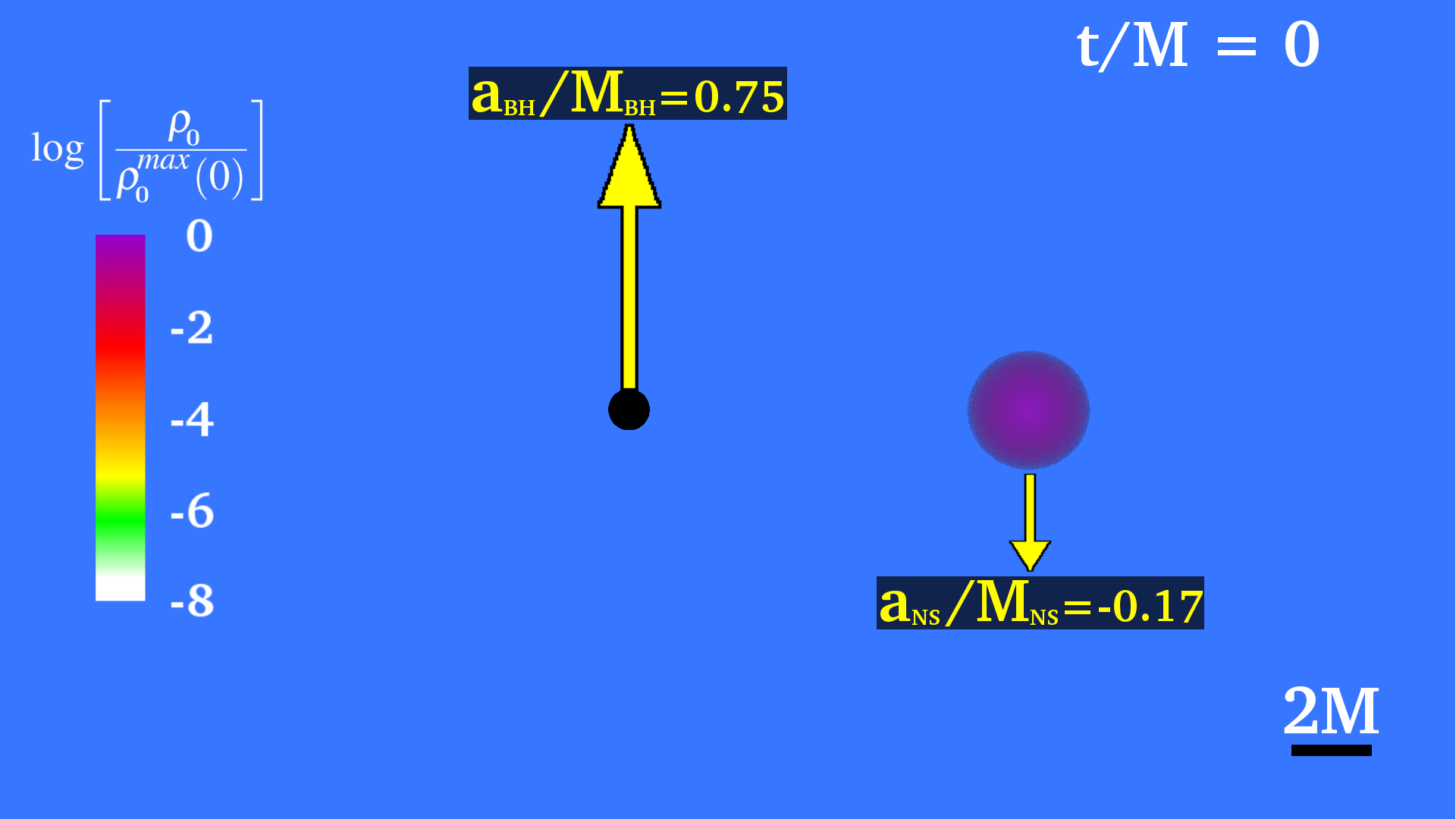}
  \includegraphics[width=0.33\textwidth]{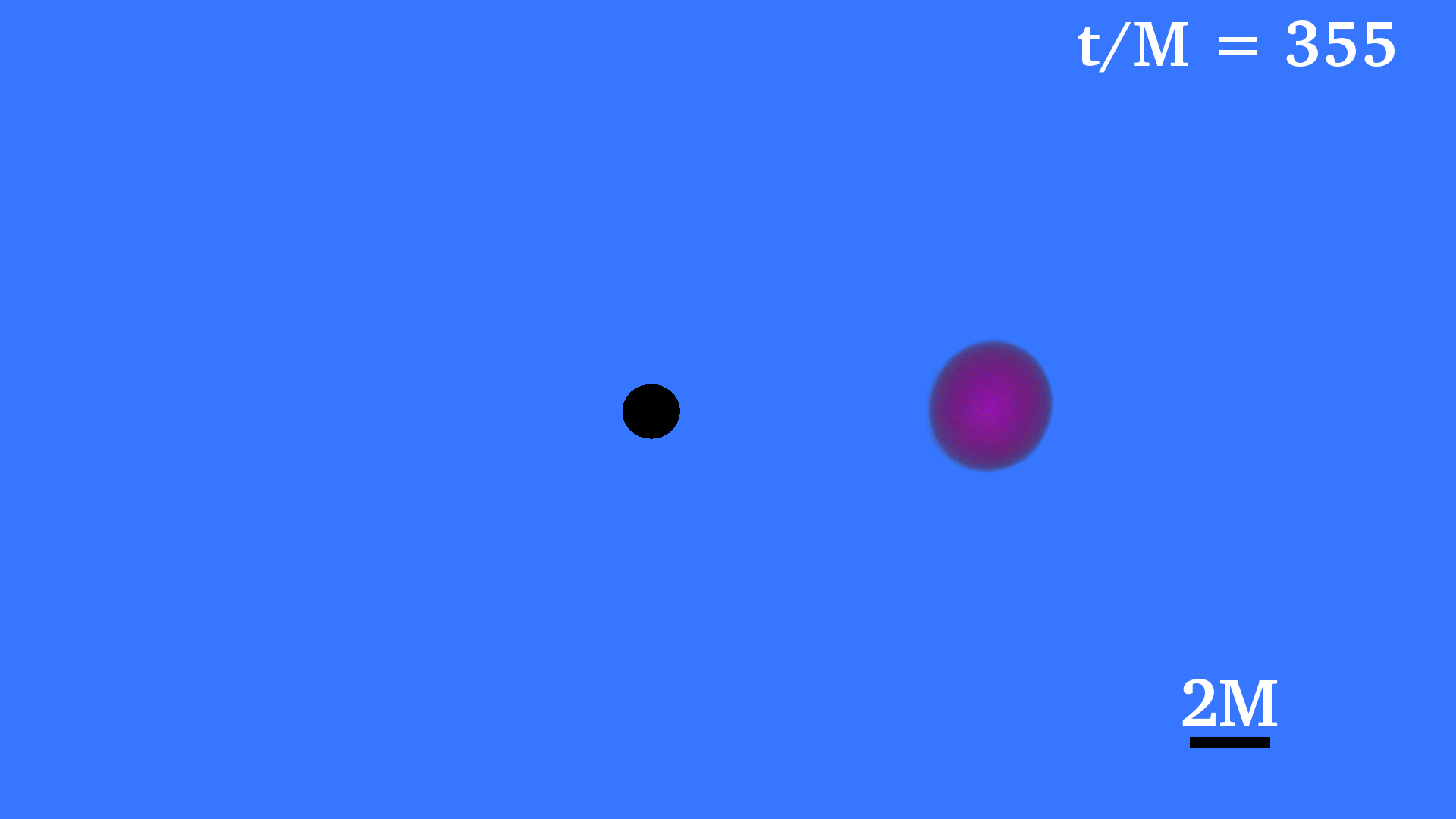}
  \includegraphics[width=0.33\textwidth]{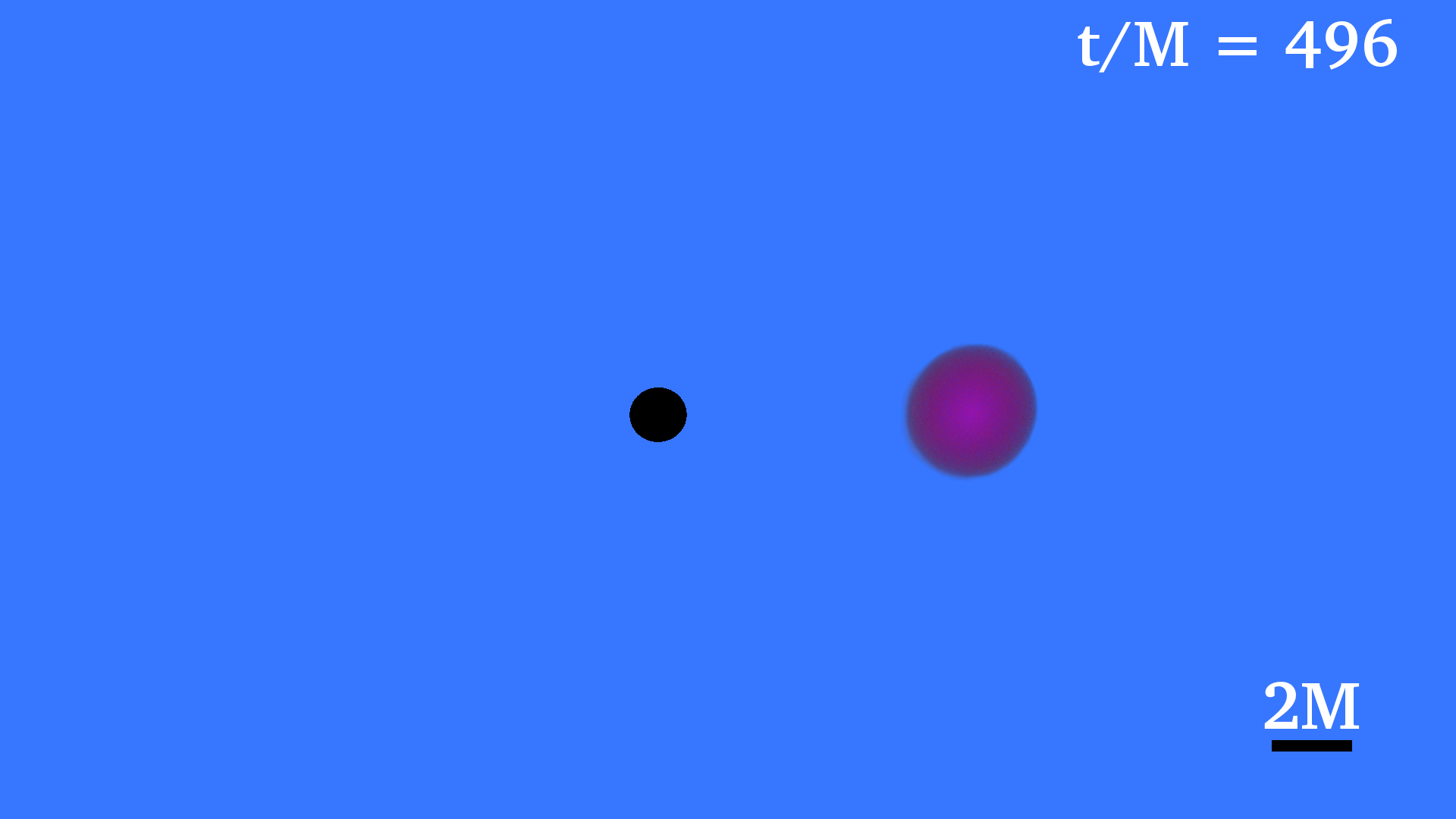}
  \includegraphics[width=0.33\textwidth]{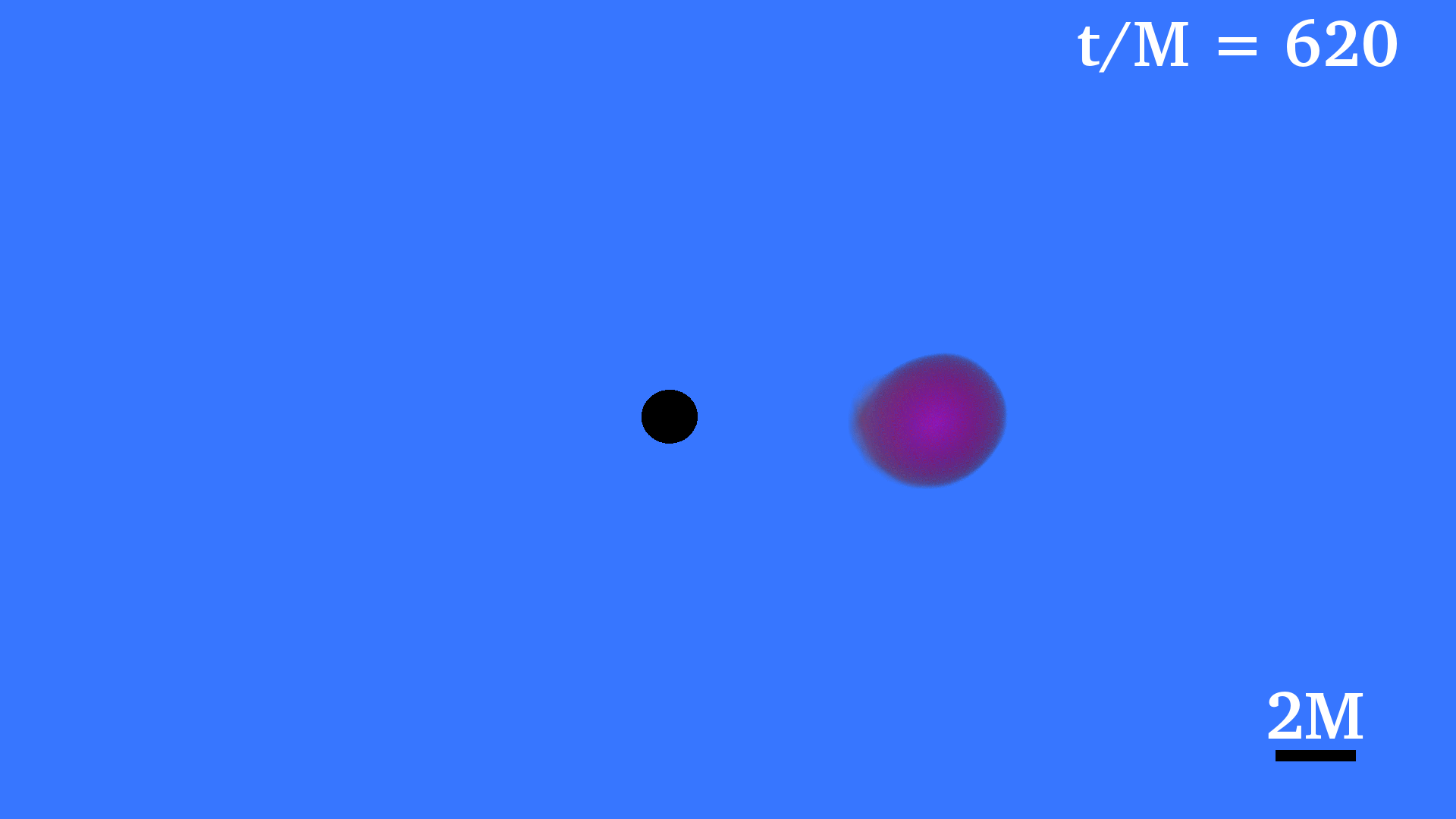}
  \includegraphics[width=0.33\textwidth]{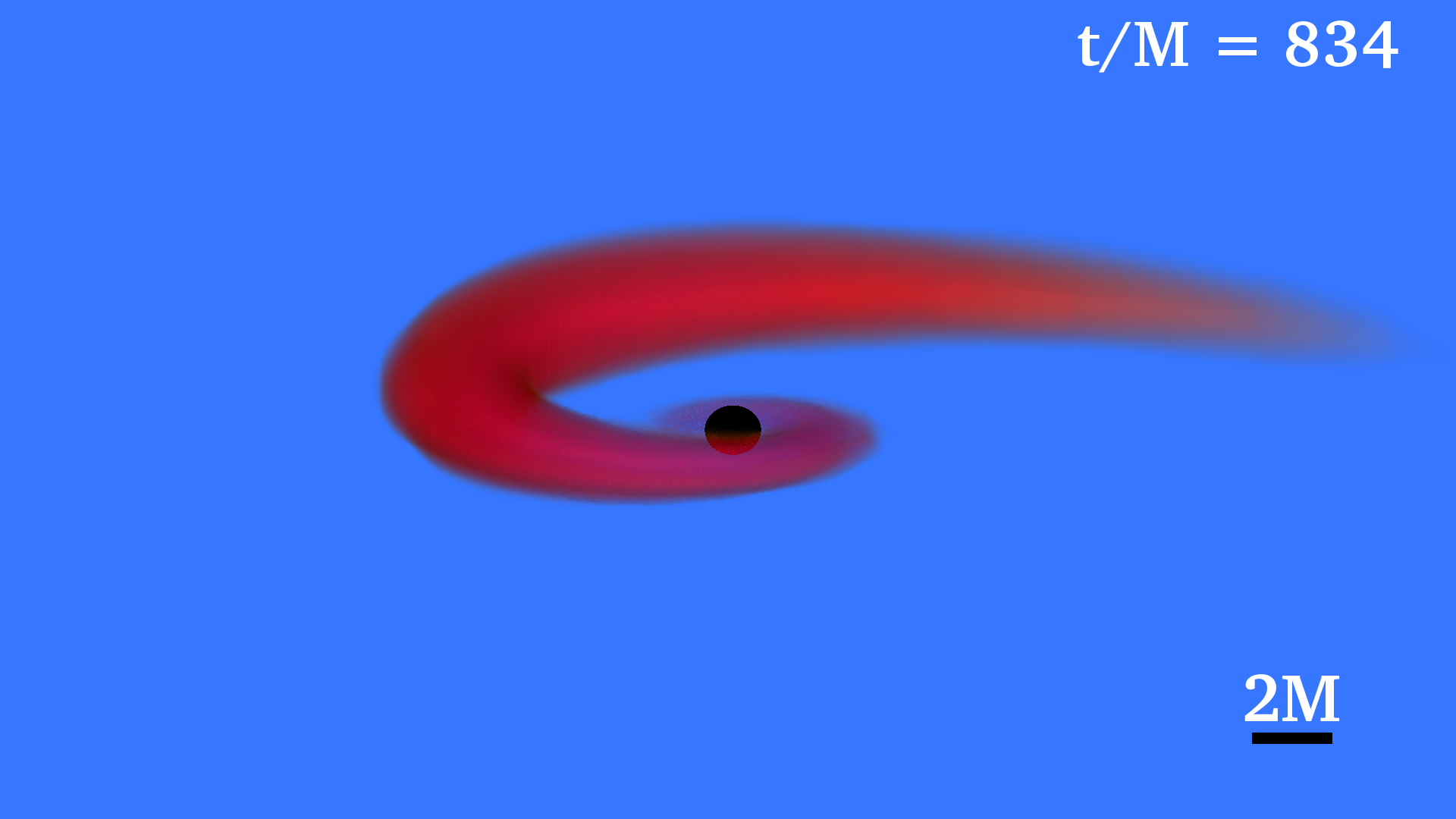}
  \includegraphics[width=0.33\textwidth]{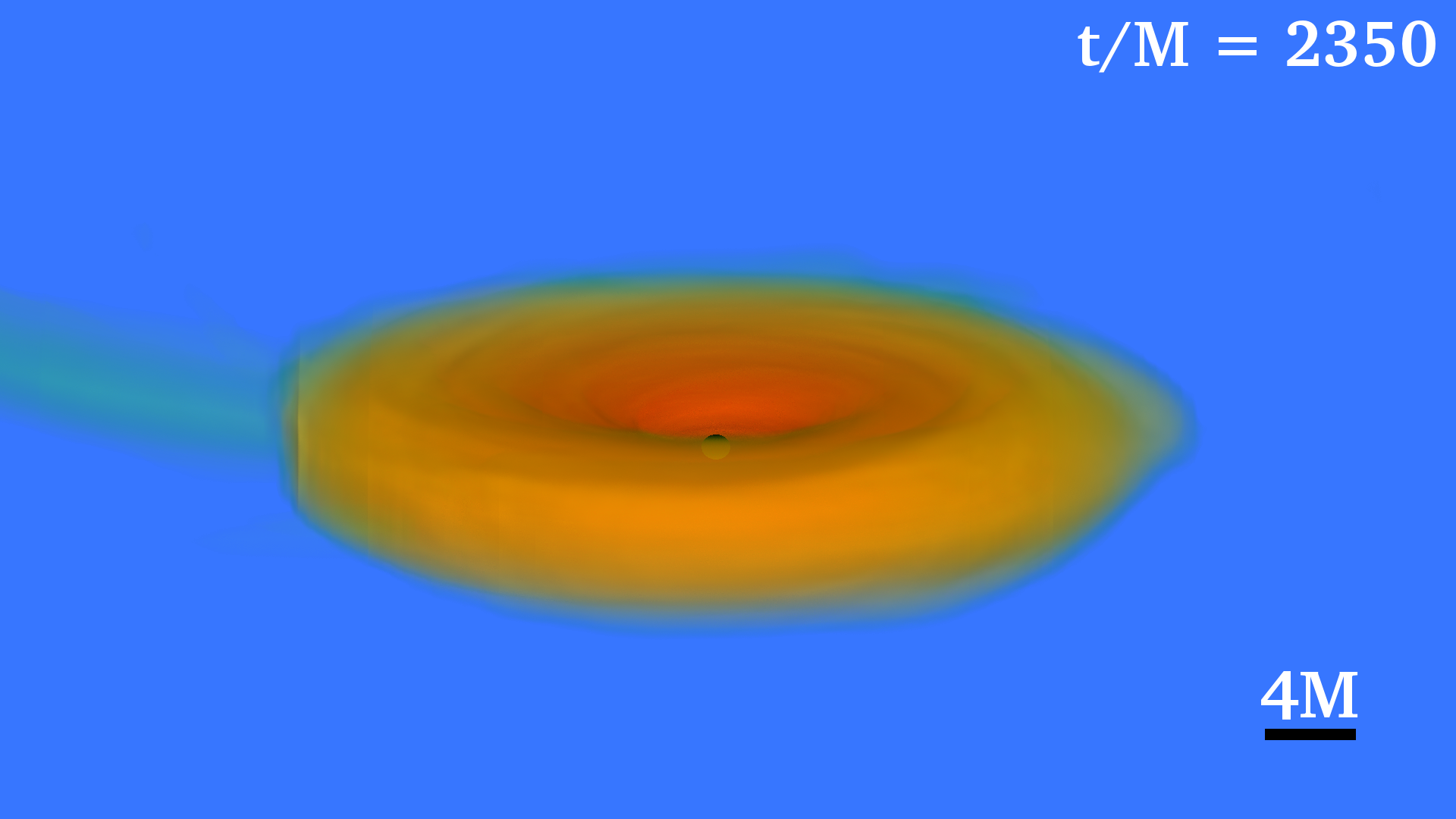}
  \caption{Volume rendering of rest-mass density $\rho_0$, normalized
    to its initial NS maximum value~$\rho_0=8.92\times
    10^{14}\,(1.4M_\odot/M_{\rm NS})^2\rm{g/cm}^{3}$ (log scale) at
    selected times for the nonmagnetized case q3NSm0.17 (see
    Table~\ref{table:BHNS_ID}). The BH apparent horizon is shown as a
    black sphere.  Top panels focus on the binary inspiral, while
    bottom ones focus on the NS tidal disruption and disk
    formation. Despite the central density oscillations, the shape of
    the NS is practically unaffected by spin, and remains nearly
    spherical during the first five of the seven orbits prior to
    merger.  Here $M=2.5\times 10^{-2}(M_{\rm NS}/1.4M_\odot){\rm
      ms}=7.58(M_{\rm NS}/1.4M_\odot)\,\rm km$.
    \label{fig:BHNS_q31:hydro}}
    \end{figure*}

\paragraph*{\bf Initial data:}
The BHNS initial data used in this work have been presented in~\cite{TBFS07b}.
The configurations
correspond to BHNS binaries on a quasicircular orbit undergoing merger with a separation chosen
to be outside the tidal disruption radius~\cite{TBFS07b}.  We consider binaries with  mass
ratio~$q=3:1$, in which the BH companion has an initial spin parameter $a_{\rm BH}/M_{\rm BH}=0.75$
($M_{\rm BH}$ is the BH Christodoulou mass~\cite{1970PhRvL..25.1596C}) aligned with the total orbital
angular momentum of the system, and binaries with mass ratio $q=5:1$ with a nonspinning (irrotational)
BH companion~(see Table~\ref{table:BHNS_ID}).
In all cases considered here, the companion has a compaction of ${\cal C}={\mathcal M}_{\rm NS}/
R_{\rm NS}=0.145$, where $\mathcal{M}_{\rm NS}$ and $R_{\rm NS}$ are the ADM mass and the
circumferential radius of the star in isolation. Note that for a polytropic EOS with
$\Gamma=2$, the maximum mass configuration has a compaction  $\mathcal{C}=0.215$.
For comparative purposes, we rescale the rest-mass of the star as 
$M_{\rm NS}=1.4M_\odot(\kappa/\kappa_L)^{1/2}$, 
and hence the maximum rest-mass density of the star is
$\rho_0^{\rm max}=8.92\times 10^{14}\,(1.4M_\odot/M_{\rm NS})^2\,\rm g/cm^{3}$. 
Here, $\kappa$ is the polytropic gas constant used to compute the initial data
and defined as $\kappa=P/\rho_0^2$, and $\kappa_L=189.96\,\rm km^2$.

Following~\cite{Ruiz:2014zta}, to induce spin we endow an irrotational
NS with an uniform angular velocity by modifying the fluid velocity as
$v^i=v^i_{\rm irrot} + {\epsilon^i}_{jk}\Omega^j\,x^k$, where
$v^i=u^i/u^0$ is the coordinate velocity of the fluid, $u^{\mu}$ is
the fluid 4-velocity, $\Omega^j$ is an angular velocity of the NS, and
${\epsilon^i}_{jk}$ is the Levi-Civita symbol.  As shown in
Table~\ref{table:BHNS_ID}, we endow the NS with spins ranging between
$-0.17$ and $0.23$ for the two BHNS mass ratios considered here. In
addition, to further assess if highly spinning NS companions can
induce potentially observable EM counterparts in BHNSs with moderate
mass ratios, we consider a more extreme case ($a_{\rm NS}/M_{\rm
  NS}=0.33$) for $q=5:1$.  In order to measure the spin of the NS, we
compute its quasilocal angular momentum $J_{\rm ql}$
\cite{Tsokaros:2019anx,Tsokaros:2018dqs}, and normalize it to the
rest-mass $J_{\rm ql}/M_{\rm NS}^2 = a_{\rm NS}/M_{\rm NS}$. If
instead we normalized by its ADM mass (in isolation) the dimensionless
spin values will increase correspondingly (e.g. our more extreme case
with spin~$a_{\rm NS}/M_{\rm NS}\sim 0.33$ becomes $\sim 0.35$). Using
the measured value of $J_{\rm ql}$, and the rest-mass $M_{\rm NS}$, we
adopt the Cook code~\cite{CookShapTeuk,Cook:1993qj,Cook:1993qr} to
generate equilibrium rotating neutron star models in isolation with
these values of angular momentum and rest-mass, and compute the
ratio of kinetic-to-gravitational potential energy $T/|W|$ as well as the
spin period $P$ that we list in Table~\ref{table:BHNS_ID}). All our
cases have an estimated $T/|W|< 0.25$, and hence the NS is stable
against the dynamical bar mode
instability~\cite{Shibata00c,BS,Paschalidis:2016vmz}. While our case
with spin $a_{\rm NS}/M_{\rm NS}=0.33$ might be unstable to the
secular $m=2$-bar mode
instability~\cite{New:2001qs,Paschalidis:2016vmz}, we point out that
the $T/|W|$ values we provide are only estimated values in isolation.

%
%
\begin{table*}[]
  \begin{center}
    \caption{Summary. Case's name starting with an NM (or an M)
      denotes a nonmagnetized (or magnetized) evolution. $a_{\rm
        BH}/M_{\rm BH}$ denotes the spin of the BH remnant, and $\Delta
      E_{\rm GW}$ and $\Delta J_{\rm GW}$ are the fraction of energy
      and angular momentum carried away by GWs, respectively. The kick
      velocity due to recoil is denoted by $v_{\rm kick}$ in $\rm
      km/s$, $M_{\rm disk}$ is the fraction of the rest-mass of the
      material outside the horizon near the end of the simulation,
      $\dot{M}$ is the rest-mass accretion rate computed via
      Eq.~(A11)~in~\cite{Farris:2009mt}, and $\tau_{\rm disk}\sim
      M_{\rm disk}/\dot{M}$ its lifetime in units of $(M_{\rm
        NS}/1.4M_\odot)\,$s, $\alpha_{\rm SS}$ is the Shakura--Sunyaev
      viscosity parameter, and $B^2/(8\pi\,\rho_0)$ is the space-averaged
      value of the force-free parameter in a cubical region of length
      $2R_{\rm BH}$ above the BH poles near the end of the
      simulation. Here $R_{\rm BH}$ is the radius of the BH
      horizon. $B_{\rm rms}$ is the rms value of the magnetic field
      above the BH poles in units of $(1.4M_\odot/M_{\rm NS})\,$G,
      $L_{\rm Poyn}$ is the Poynting luminosity driven by the jet in
      units of $\rm erg/s$ and time-averaged over the last~$500M\sim
      12.5(M_{\rm NS}/1.4M_\odot)\rm ms$ of the evolution, $v_{\rm
        eje}$ and $M_{\rm eje}$ are the mass-averaged velocity and
      rest-mass of the unbound material, while $L_{\rm knova}$ and
      $t_{\rm peak}$ are the peak luminosity and the rise time of the
      potential kilonova in units of $\rm erg/s$ and days,
      respectively. A dash symbol denotes no corresponding/unavailable
      value.
      \label{table:summary_BHNSresults}}
    \begin{adjustbox}{width=\linewidth,center}
    \begin{tabular}{ccccccccccccccccc}
      \hline\hline
          {Case} &$a_{\rm BH}/M_{\rm BH}$&$\Delta E_{\rm GW}/M$ &$\Delta J_{\rm GW}/J$&
          $v_{\rm kick}$ &$M_{\rm disk}/{M_{\rm NS}}$ & $\dot{M} (M_\odot/s)$ &$\tau_{\rm disk}$ &$\alpha_{\rm SS}$ & $B^2/(8\pi\,\rho_0)$ &$B_{\rm rms}$ &$L_{\rm Poyn}$ & $v_{\rm eje}$
          & $M_{\rm eje}$
          &$L_{\rm knova}$ & $t_{\rm peak}$\\
          \hline
          \hline
          NMq3NS0.0$^\dag$     & 0.89  & $0.97\%$ & $14.89\%$ & $54.20$ & $10.90\%$     & $0.39$ & $0.4$  &$-$ &$-$ &$-$ &$-$ & $-$ &$-$ &$-$&$-$\\
          NMq3NSm0.17   & 0.89  & $1.14\%$ & $13.10\%$ & $44.62$ & $10.84\%$     & $0.46$ & $0.3$  &$-$ &$-$ &$-$ &$-$ &0.24 &$10^{-2.4}$ &$10^{41.2}$ &0.18\\
          NMq3NSp0.23   & 0.90  & $1.10\%$ & $14.48\%$ & $63.96$ & $14.56\%$     & $0.28$ & $0.7$  &$-$ &$-$ &$-$ &$-$ & 0.29&$10^{-2.3}$ &$10^{41.2}$ &0.18\\
          \hline
          Mq3NS0.0$^\dag$     & 0.85  & $1.11\%$ & $14.91\%$ & $54.20$ & $10.00\%$     & $0.25$ & $0.5$  & $0.01-0.03$   &  $10^{2.0}$   & $10^{15.2}$  & $10^{51.2}$ & $-$&$-$ &$-$&$-$\\
          Mq3NSm0.17   & 0.85  & $1.20\%$ & $12.90\%$ & $46.40$ & $8.78\%$      & $0.29$ & $0.5$  & $0.01-0.03$   &  $10^{2.0}$   & $10^{15.1}$  & $10^{51.6}$ &0.25 & $10^{-2.1}$&$10^{41.3}$ &0.25\\
          Mq3NSp0.23   & 0.87  & $1.12\%$ & $14.60\%$ & $64.81$ & $14.17\%$     & $0.23$ & $0.8$  & $0.01-0.03$   &  $10^{2.1}$   & $10^{15.2}$  & $10^{52.1}$ &0.27 & $10^{-2.0}$&$10^{41.4}$ &0.27\\         
          \hline\hline
          NMq5NS0.0$^\dag$     & 0.42  & $1.07\%$ & $20.07\%$ & $67.96$ & $0.29\%$     & $0.02$ & $0.2$ &$-$ &$-$ &$-$ &$-$ &$-$ &$-$   &$-$\\
          NMq5NSm0.17   & 0.42  & $1.00\%$ & $19.56\%$ & $34.74$ & $0.28\%$     & $0.02$ & $0.2$ &$-$ &$-$ &$-$ &$-$ & 0.26 &$10^{-4.7}$&$10^{40.0}$ & 0.01\\
          NMq5NSp0.23   & 0.42  & $1.08\%$ & $20.29\%$ & $62.88$ & $0.65\%$     & $0.12$ & $0.1$ &$-$ &$-$ &$-$ &$-$ & 0.27 &$10^{-3.7}$&$10^{40.4}$ & 0.02\\
          NMq5NSp0.33   & 0.43  & $1.07\%$ & $20.35\%$ & $91.10$ & $1.23\%$     & $0.15$ & $0.1$ &$-$ &$-$ &$-$ &$-$ & 0.33 &$10^{-3.5}$&$10^{40.6}$ & 0.04\\
           \hline 
          Mq5NS0.0$^\dag$     & 0.42  & $1.05\%$ & $19.63\%$ & $69.96$ & $0.53\%$     & $0.04$ & $0.2$  & $-$            &  $10^{-3.0}$  & $10^{12.3}$  & $-$ &$-$ &$-$ &$-$&$-$ \\
          Mq5NSm0.17   & 0.42  & $1.00\%$ & $19.43\%$ & $36.82$ & $0.53\%$     & $0.02$ & $0.3$  & $-$            &  $10^{-3.2}$  & $10^{12.4}$  & $-$ &0.25 &$10^{-4.7}$ &$10^{40.0}$ & 0.01  \\
          Mq5NSp0.23   & 0.42  & $1.06\%$ & $20.14\%$ & $70.31$ & $1.04\%$     & $0.10$ & $0.1$  & $-$            &  $10^{-3.1}$  & $10^{12.1}$  & $-$ &0.27 &$10^{-3.7}$ &$10^{40.4}$ & 0.03 \\
          Mq5NSp0.33   & 0.43  & $1.06\%$ & $20.20\%$ & $93.80$ & $1.42\%$     & $0.12$ & $0.2$  & $-$            &  $10^{-2.9}$  & $10^{12.7}$  & $-$ &0.35 &$10^{-3.5}$ &$10^{40.6}$ & 0.04 \\
          \hline\hline
    \end{tabular}
    \end{adjustbox}
    \begin{flushleft}
   $^{\dag}$ Cases treated previously in~\cite{Ruiz:2018wah}.
  \end{flushleft}
  \end{center}
\end{table*} 

As the new fluid velocity field no longer satisfies the hydrostatic
equations, the NS undergoes small radial
oscillations. Fig.~\ref{fig:rho_spin} shows the relative changes in
the central rest-mass density during the early inspiral for all cases
in Table~\ref{table:BHNS_ID}.  In contrast to the quasiequilibrium
(irrotational) case, where the amplitude of the oscillations stays
below~$\sim 1\%$, the amplitude of the oscillations in cases $a_{\rm
  NS}/M_{\rm NS}=-0.17$ and $0.23$ (see Table~\ref{table:BHNS_ID}) is
larger but remains below $\sim 8\%$, and below~$\sim 16\%$ in our
extreme case ($a_{\rm NS}/M_{\rm NS}=0.33$).  The inset shows the
$L_2$ norm of the Hamiltonian constraint for cases with mass ratio
$q=5:1$ during the early inspiral. We observe that due to the
constraint damping~(see Eq.~19 in~\cite{DMSB}) used in our numerical
evolutions, the constraint violation induced by our {\it ad-hoc}
prescription of the new fluid velocity is damped and propagated away
after roughly one orbit (or $t\lesssim 220M$). Similar results are
observed on the other cases in Table~\ref{table:BHNS_ID}.

Next, we evolve the above configurations until about two orbits before
tidal disruption. At that point, the NS is threaded by a dynamically
weak, dipolar magnetic field induced by a vector potential generated
by a current loop inside the star (see
Eq.~2~in~\cite{Paschalidis:2013jsa}).  As
in~\cite{prs15,Ruiz:2018wah}, we choose the current $I_0$ and the
radius of the loop $r_0$ such that the magnetic-to-gas-pressure ratio
at the center of the NS is $P_{\rm mag}/P_{\rm gas}=10^{-2.5}$
(see~Fig.~2~in~\cite{Ruiz:2017inq}).  The resulting magnetic field at
the pole of the star turns out to be $B_{\rm pole}\sim 6\times 10^{15}
(1.4M_\odot/M_{\rm NS})\,\,$G.  As pointed out in~\cite{prs15},
although we choose an astrophysically large magnetic field, it is
dynamically unimportant in the stellar interior and does not affect
the late inspiral or the merger phases. We do expect that the outcome
of our numerical results will apply to other dynamically weak field
choices because the magnetic field amplification following merger 
will be mainly triggered by magnetic winding and the
magneto-rotational-instability (MRI)~\cite{kskstw15}.

On the other hand, to capture one of the properties of the force-free
conditions that likely characterize the NS exterior (magnetic-pressure
dominance), we set a variable and low-density magnetosphere outside
the star such that the magnetic-to-gas pressure ratio
is~$\beta^{-1}=P_{\rm gas}/ P_{\rm mag}=0.01$
everywhere~\cite{prs15}. This one-time reset of the low-density
magnetosphere increases the total rest-mass on the entire grid by less
than~$1\%$.
%
\paragraph*{\bf Grid structure:}
The grid hierarchies used to evolve the BHNS binaries with mass ratio
$q=3:1$ and $q=5:1$ are the same as those used to evolve models
Tilq3sp0.75 and Aliq5sp0.0 (see~Table~II of~\cite{Ruiz:2018wah}),
respectively. They consist of two sets of nested refinement levels
centered on both the BH and the NS. In all cases, the NS is covered by
eight refinement levels, while the BH companion is covered by nine
refinement levels for $q=3:1$ cases, or by eight levels for $q=5:1$
cases. The finest resolution box has a half length of $\sim
1.5\,R_{\rm BH}$ around the BH and $\sim 1.2\,R_{\rm NS}$ around the
NS. These choices resolve the initial equatorial radius of the BH
apparent horizon by around $40$ grid points, and the initial NS
equatorial radius by around $42$ grid points. In all cases we impose
reflection symmetry across the orbital plane ($z=0$).

%
\section{Results}
\label{sec:results}
\begin{figure*}
  \centering
  \includegraphics[width=0.33\textwidth]{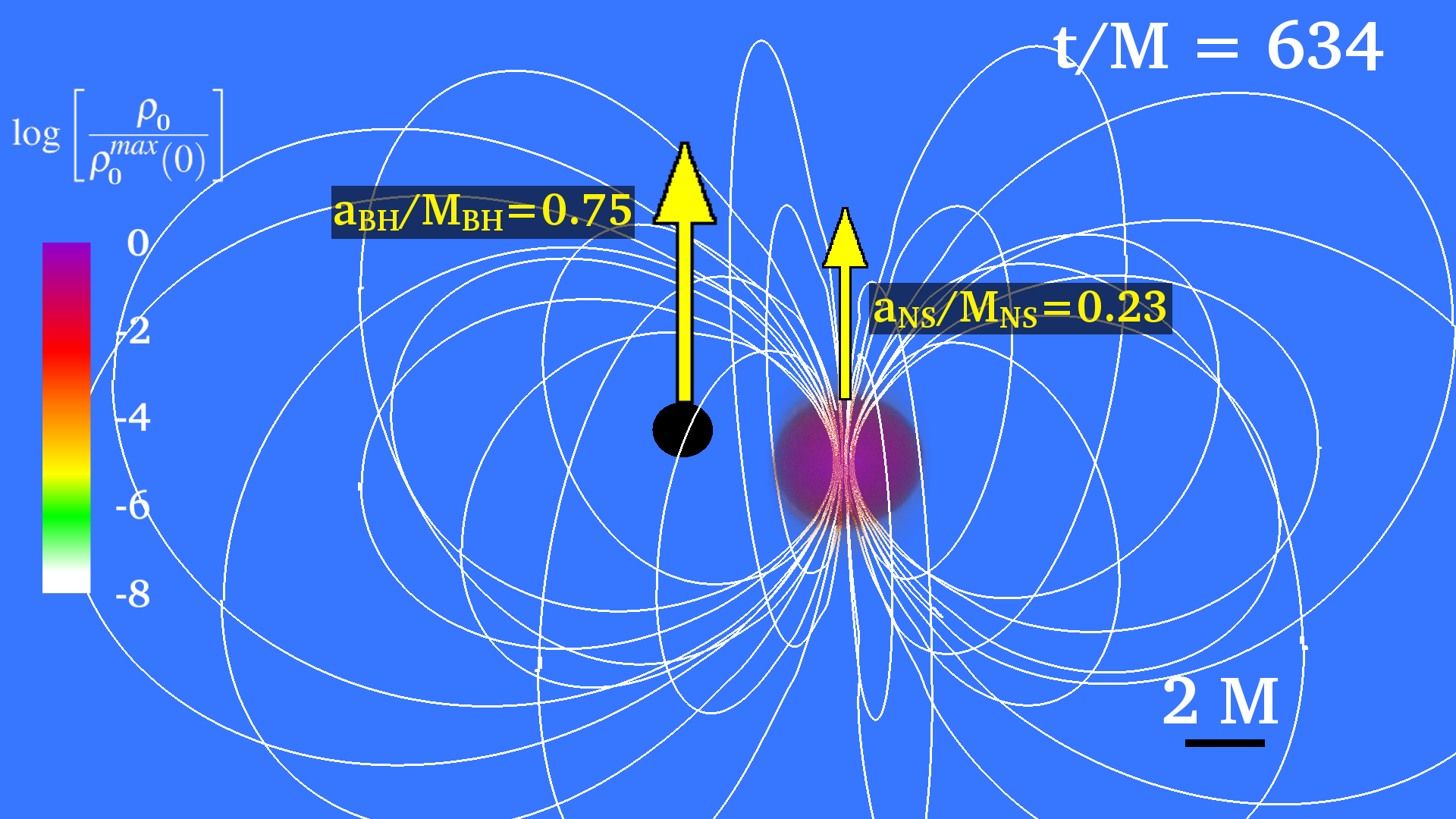}
  \includegraphics[width=0.33\textwidth]{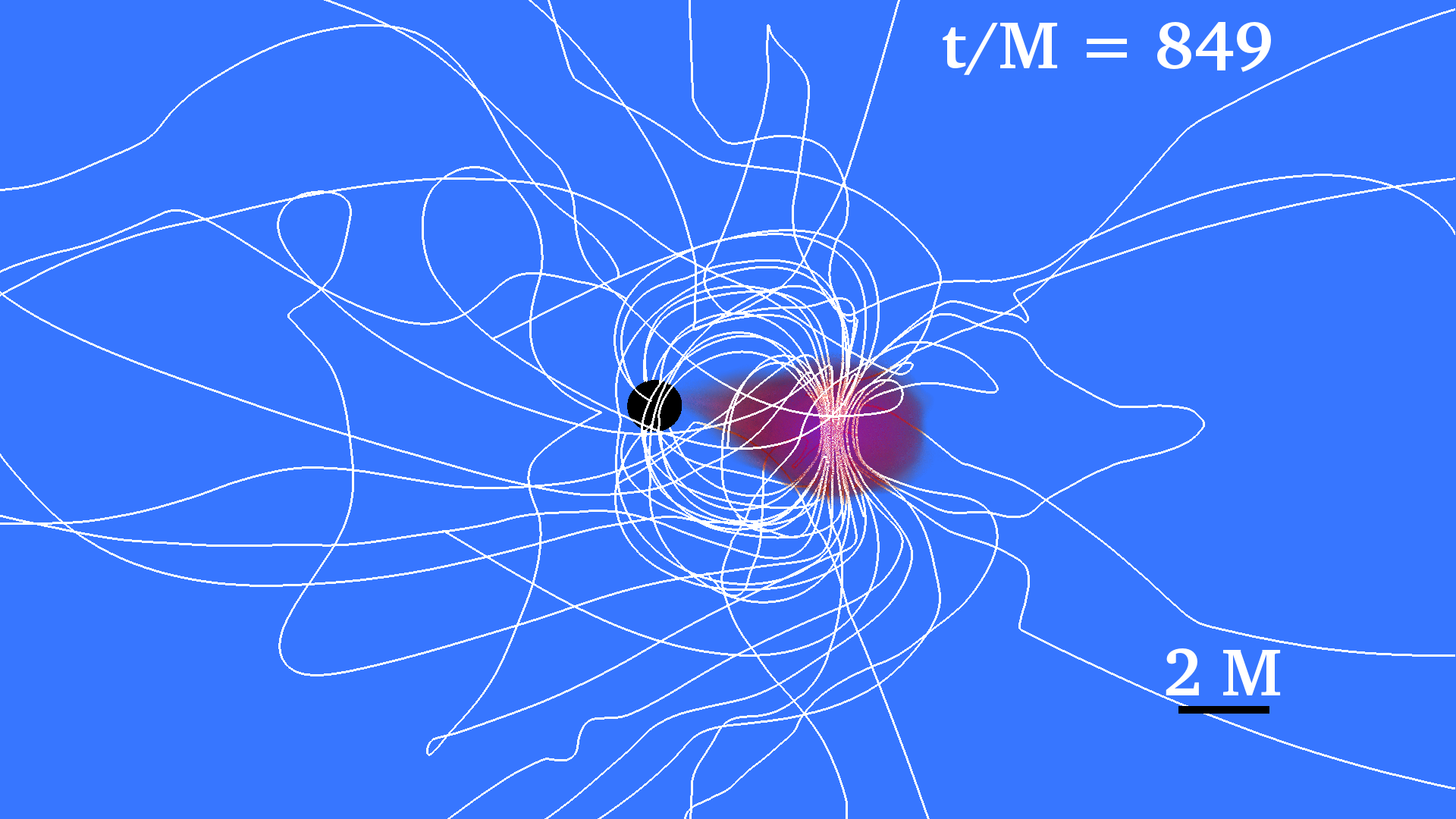}
  \includegraphics[width=0.33\textwidth]{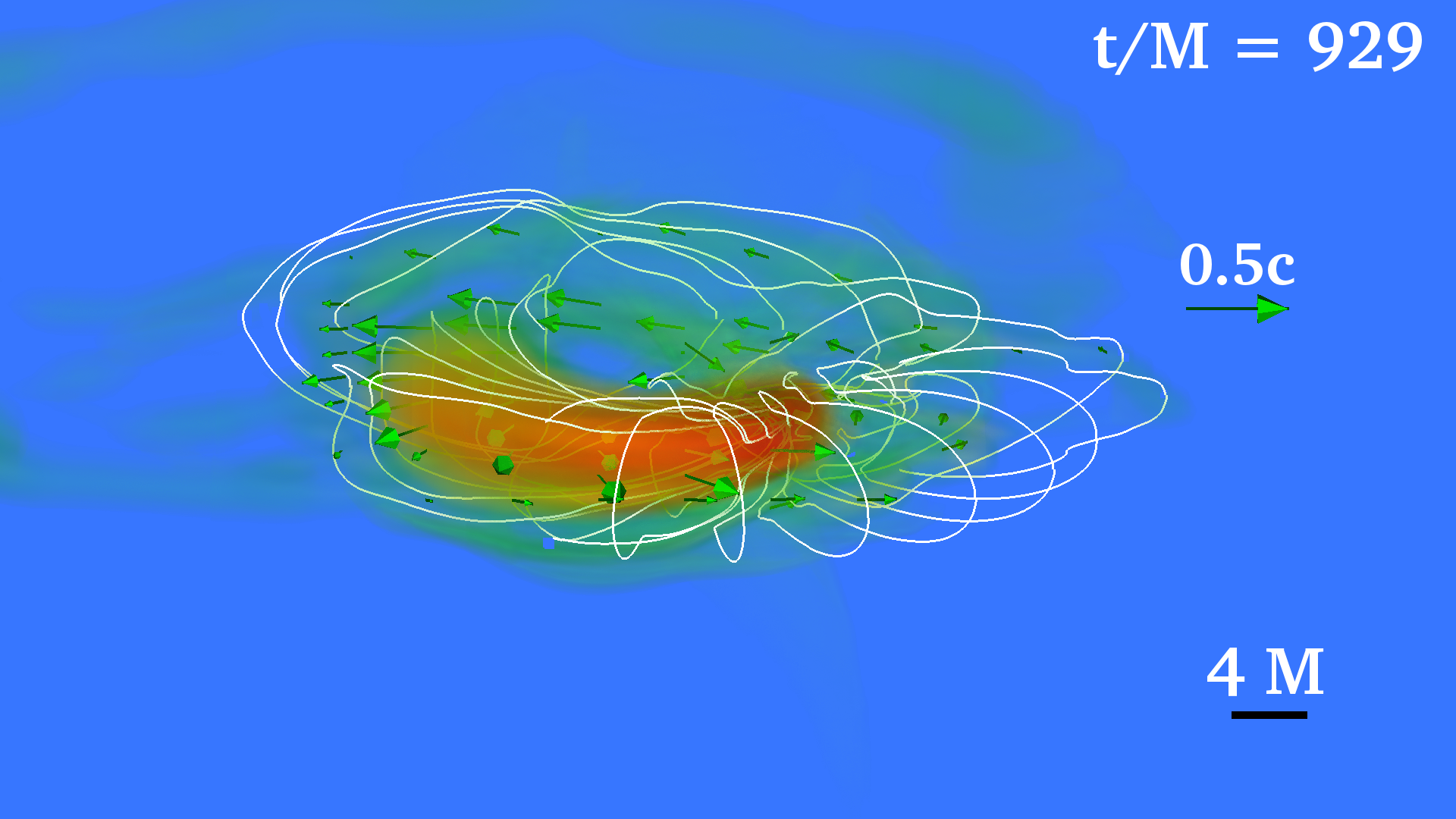}
  \includegraphics[width=0.33\textwidth]{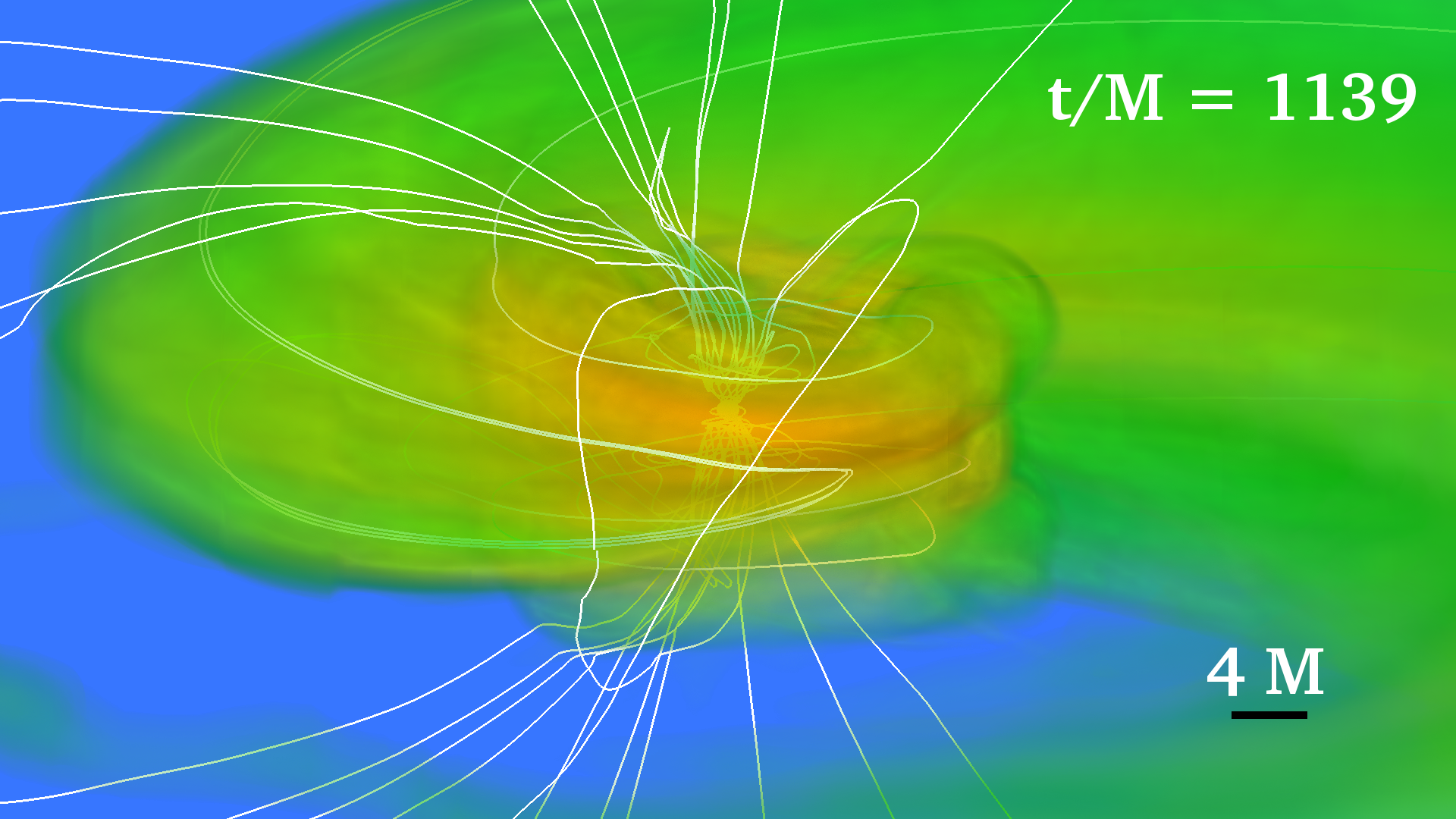}
  \includegraphics[width=0.33\textwidth]{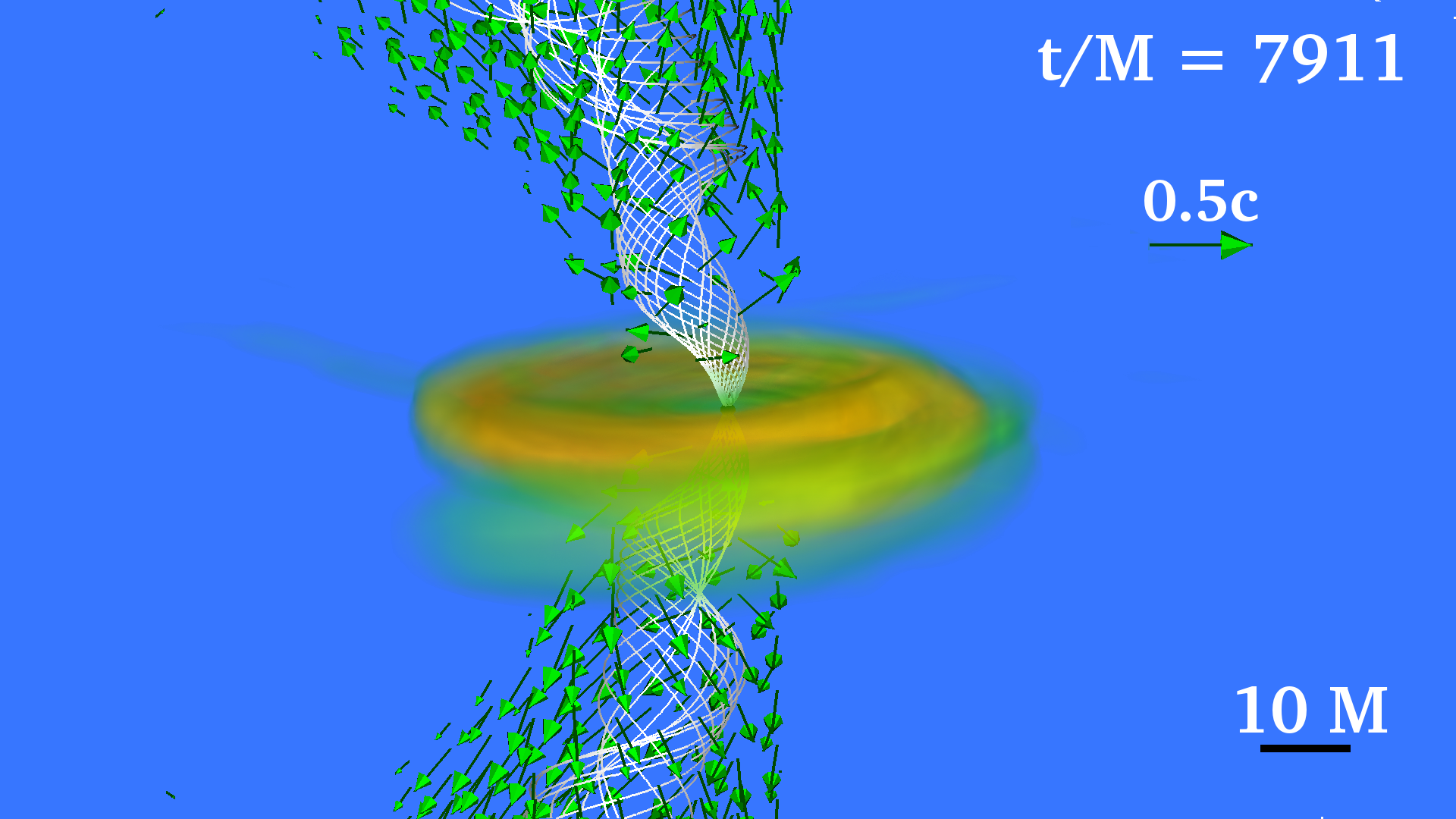}
  \includegraphics[width=0.33\textwidth]{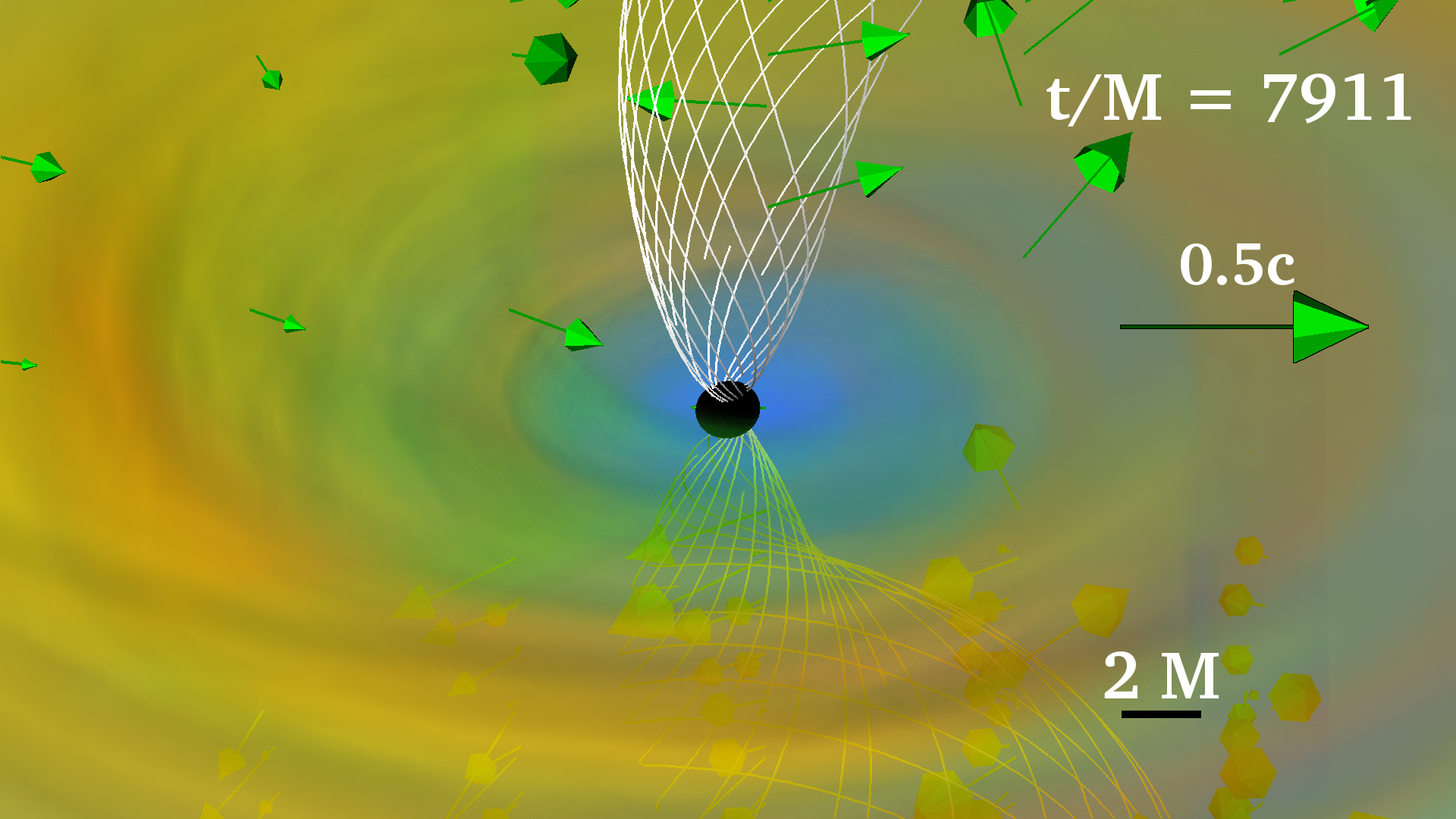}
\caption{Volume rendering of rest-mass density $\rho_0$ normalized to its initial NS
    maximum value~$\rho_0=8.92\times 10^{14}\,(1.4M_\odot/M_{\rm NS})^2\rm{g/cm}^{3}$
    (log scale) at selected times for Mq3NSp0.23 (see Table~\ref{table:BHNS_ID}).
    Bottom panels highlight the emergence of the magnetically-driven jet.
    White lines denote the magnetic field, arrows denote the fluid velocity, while the
    BH apparent horizon is shown as a black sphere.  Here $M=2.5\times 10^{-2}(M_{\rm NS}/
    1.4M_\odot)\,{\rm ms}=7.58(M_{\rm NS}/1.4M_\odot)\,\rm km$.
    \label{fig:BHNS_case_q31_OmegaH}}
\end{figure*}

The basic dynamics and outcomes of irrotational NS cases in Table~\ref{table:summary_BHNSresults}
have been previously described in~\cite{Etienne:2008re,prs15,Ruiz:2018wah}. There it was found that,
in contrast to the high mass ratio cases where the NS basically plunges into the BH,  the NS
in~$q=3:1$~cases is tidally disrupted before reaching the ISCO. This resulted in long tidal tails
of matter that eventually settle down, forming a significant accretion disk around the BH regardless
of the magnetic field content~(see Figs.~\ref{fig:BHNS_q31:hydro}~and~\ref{fig:BHNS_case_q31_OmegaH}).
%

In the magnetized cases, matter wrapping around the BH drags the
frozen-in magnetic field into a predominantly toroidal
configuration. However, initially the magnetic field lines connect
the star with the poles of the BH, and later the low-density debris
ejected during the tidal disruption remains connected to the
accretion disk via these field lines~(see top panels in
Figs.~\ref{fig:BHNS_case_q31_OmegaH} and~\ref{fig:BHNS_case_s0_sm05}),
as a result, the external magnetic field maintains a strong poloidal
component. Magnetic winding and the MRI then amplify the magnetic
field above the BH poles from $\sim 10^{13}(1.4M_\odot/M_{\rm NS})\,$G
to $\sim 10^{15} (1.4M_\odot/M_{\rm NS})\,$G, when the accretion disk
settles down. This amplification induces high magnetic pressure
gradients above the BH poles that, when the regions above the BH poles
approach force-free values ($B^2/8\,\pi \rho_0\gg 1$), lead to the
launching of a mildly relativistic outflow with a Lorentz factor
$\Gamma_L\gtrsim 1.2$ confined inside a tightly wound, helical
magnetic field funnel --an incipient jet~(see bottom panels in~Figs.
\ref{fig:BHNS_case_q31_OmegaH} and~\ref{fig:BHNS_case_s0_sm05}). Here
$B$ and $\rho_0$ are the strength of the magnetic field and the
rest-mass density, respectively.  The lifetime of the jet and its
associated luminosity are consistent with typical
sGRBs~\cite{Bhat:2016odd,
  Lien:2016zny,Svinkin:2016fho,Ajello:2019zki}, as well as with the BZ
mechanism~\cite{BZeffect,Shapiro:2017cny}.
%
\subsection{Mass ratio $q=3:1$}
\label{sec:q=3_sec}
Figure~\ref{fig:BHNS_q31:hydro} shows snapshots of the rest-mass
density at selected times for NMq3NSm0.17~(see
Table~\ref{table:summary_BHNSresults} for an explanation of the
augmented case labels). We observe that although the star undergoes
small radial oscillations due to our {\it ad hoc} prescription for the
NS spin~(see~Fig.~\ref{fig:rho_spin}), the shape of the star is nearly
spherical during the first five of the almost seven inspiral orbits
before the binary merger~(see top panels in
Fig.~\ref{fig:BHNS_q31:hydro}).  Bottom panels focus on the NS tail
deformation, merger, and the subsequent formation of a quasistationary
disk, as matter having larger specific angular momentum wraps around
the BH. Similar behavior is observed in the other cases independent of
the magnetic field (see
Table~\ref{table:summary_BHNSresults}). However, we observe that the
larger the NS prograde spin is, the further out the NS disruption take
place.

Notice that as the BH spin increases the ISCO decreases, and hence the
tidal disruption effects become more pronounced, resulting in long
tidal tails that eventually wrap around the BH forming the accretion
disk. Similarly, as the prograde NS spin
increases, the effective ISCO decreases~(see
e.g.~\cite{Barausse:2009xi}). Additionally, as the magnitude of the NS
spin increases, the star becomes less bound, and the tidal separation
radius $r_{\rm tid}$ (the separation at which tidal disruption of the
NS begins) increases, resulting also in more pronounced disruption
effects. This effect can be easily understood by estimating~$r_{\rm
  tid}$ through a simple Newtonian argument. Equating the inward
gravitational force exerted by the NS on its fluid elements with the
BH's outgoing tidal force and the outgoing centrifugal force we find
that (see also~\cite{East:2015yea} for a similar expression)
\begin{equation}
  r_{\rm tid}/M_{\rm BH}\simeq q^{-2/3}\,\mathcal{C}^{-1}\,
  \left[1-\Omega^2\,M_{\rm NS}^2\,\mathcal{C}^{-3}\right]^{-1/3}\,,
 \label{eq:tidal}
\end{equation}
where $\Omega=a_{\rm NS}M_{\rm NS}/I$. Here $I$ is the moment of
inertia of the star. Therefore, the larger the magnitude of the NS
spin, the larger $r_{\rm tid}$, and hence the more material 
spreads out to form the disk.  Consistent with the above predictions,
we find that the accretion disk in our extreme cases has a rest mass
ranging between~$\sim 9\%$ (for Mq3NSm0.17) and $\sim 14.2\%$~(for
Mq3NSp0.23)~of the total rest-mass of the NS~(see~top panel~in
Fig.~\ref{fig:M0_outside}). Slightly more massive disks are found in
the~nonmagnetized~cases~(see~Table~\ref{table:summary_BHNSresults}).
%
\begin{figure}
  \centering
  \includegraphics[width=0.45\textwidth]{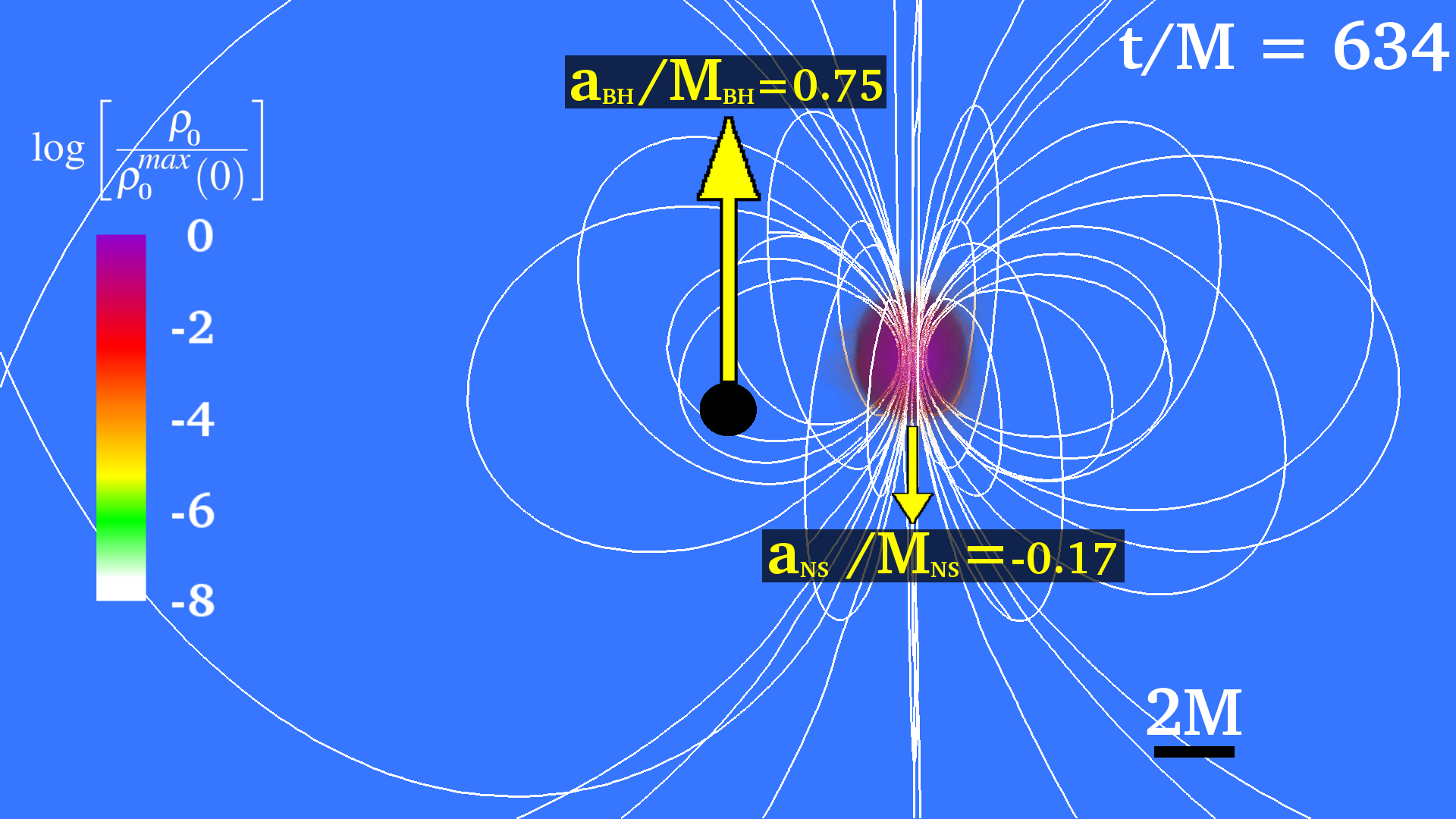}
  \includegraphics[width=0.45\textwidth]{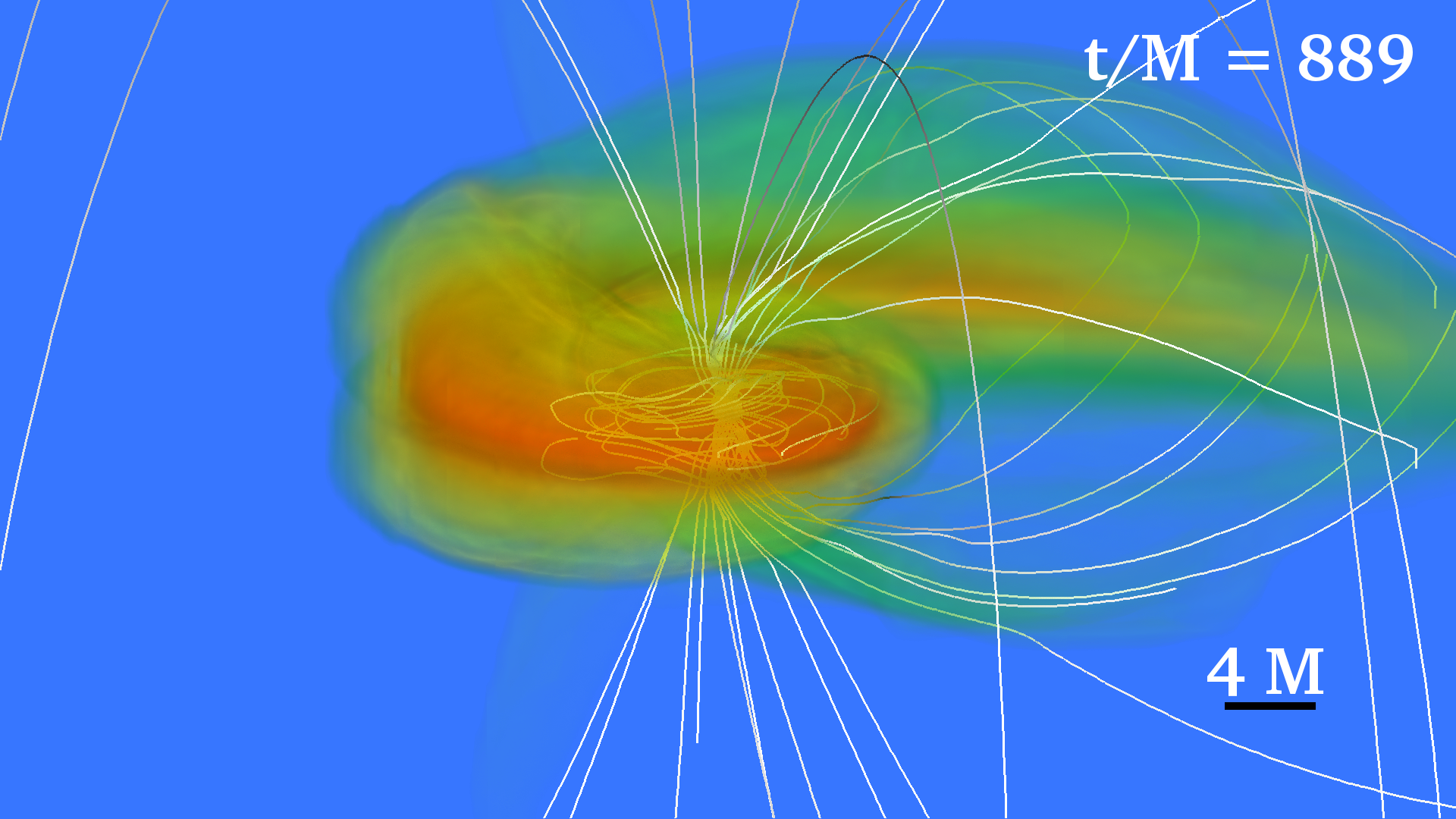}
  \includegraphics[width=0.45\textwidth]{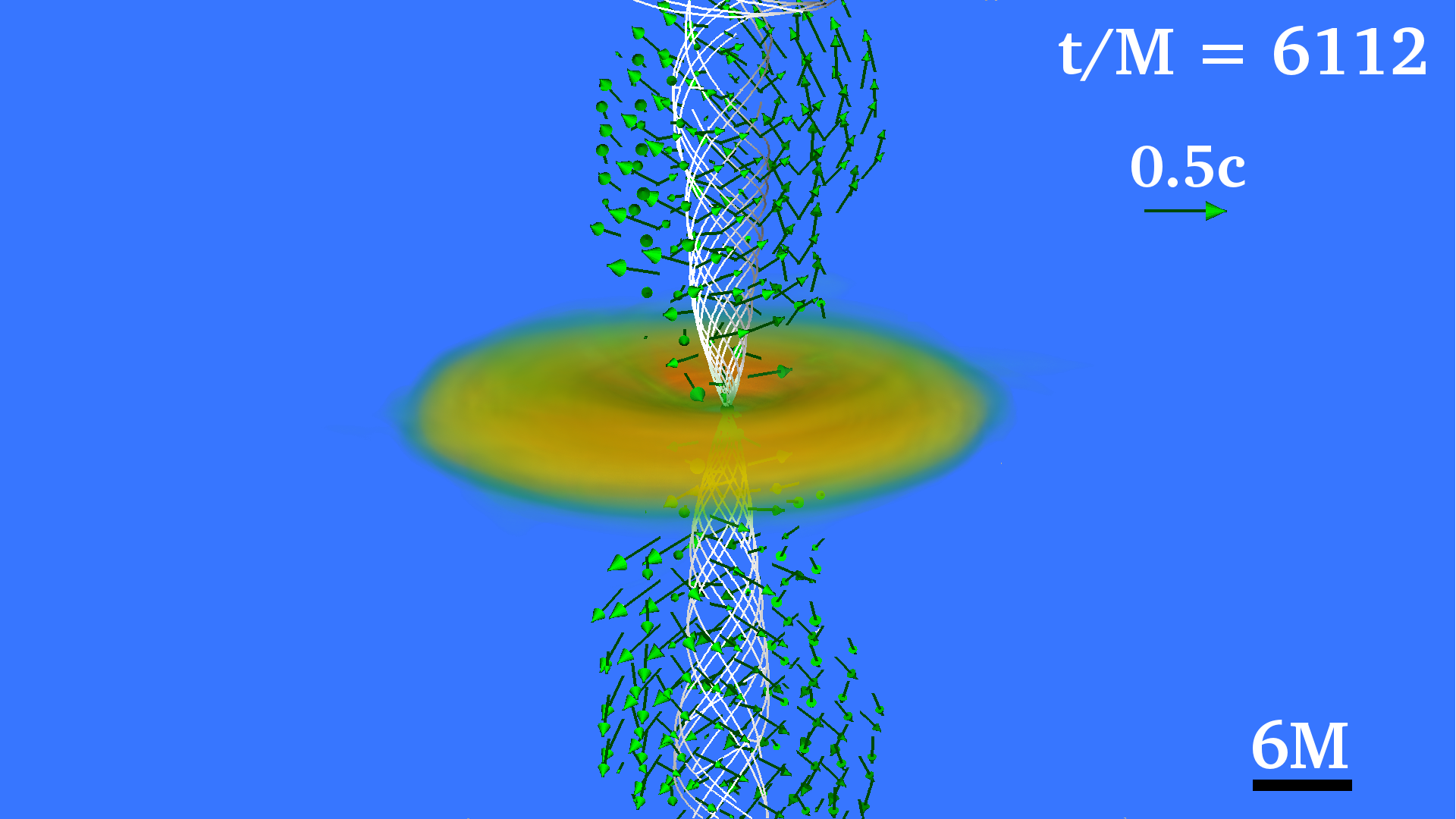}
  \caption{Same as Fig.~\ref{fig:BHNS_case_q31_OmegaH} but for 
    Mq3NSm0.17.\label{fig:BHNS_case_s0_sm05}}
\end{figure}
In all cases, the BH remnant has a mass of~$\simeq 4.76M_\odot(M_{\rm
  NS}/1.4M_\odot)$, and its spin is $a_{\rm BH}/M_{\rm BH}\simeq 0.9$
for the nonmagnetized cases, and $a_{\rm BH}/M_{\rm BH}\simeq 0.85$
for the magnetized cases. These values seem to be unaffected by the
initial NS spin. Similar behavior was reported in spinning BNS
mergers~\cite{Ruiz:2019ezy}.

By contrast to the nonmagnetized cases, where the BH + disk remnant settles down into an
almost steady configuration after $\sim 800M\simeq 20(M_{\rm NS}/1.4M_\odot)\,
\rm ms$ following merger~(see~bottom panels of~Fig.~\ref{fig:BHNS_q31:hydro}), the
magnetized cases launch a mildly relativistic outflow confined in a tightly wound, helical
magnetic field funnel after~$\sim 3500-5500M\simeq 88-138(M_{\rm NS}/1.4M_\odot)
\,\rm ms$ (see below) following merger (see bottom panels in~Figs.
\ref{fig:BHNS_case_q31_OmegaH} and~\ref{fig:BHNS_case_s0_sm05}).

\begin{figure}
  \centering
  \includegraphics[width=0.48\textwidth]{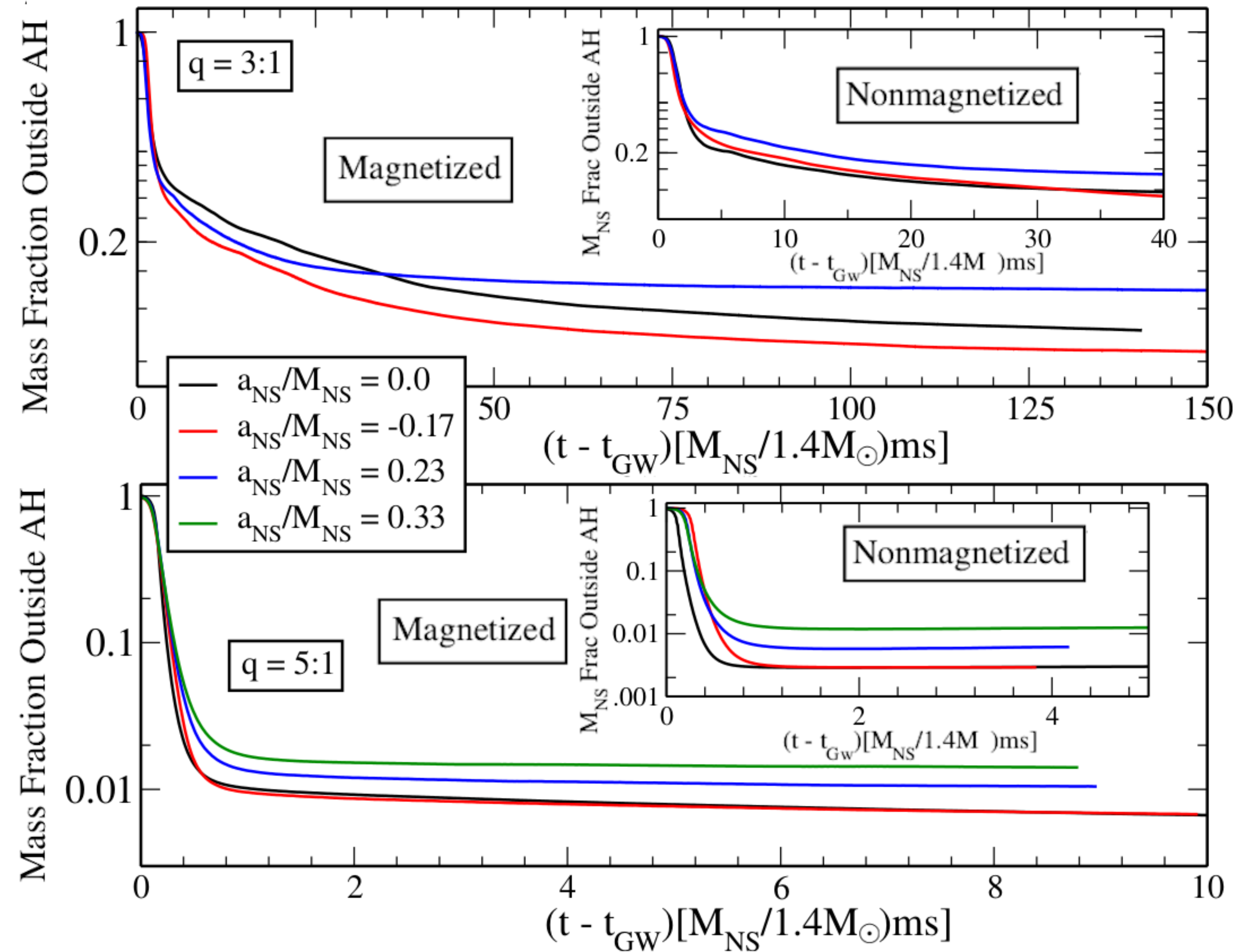}
  \caption{rest-mass fraction outside the BH apparent horizon as a function of the coordinate time
    for cases listed in Table~\ref{table:summary_BHNSresults}. The insets focus on pure
    nonmagnetized cases. The coordinate time has been shifted to the merger time $t_{\rm mer}$. 
    \label{fig:M0_outside}}
\end{figure}

To probe if magnetic turbulence is operating within the accretion
disk, we first verify that the wavelength of the fastest growing MRI
mode~$\lambda_{\rm MRI}$~in all our cases is resolved by $\gtrsim 5$
grid points (see Fig.~9~in~\cite{Ruiz:2018wah}). It is likely that the
MRI is at least partially captured in our
simulations~\cite{Gold:2013zma}. We also compute the Shakura--Sunyaev
viscosity $\alpha_{\rm SS}$~parameter through
Eq.~26~in~\cite{FASTEST_GROWING_MRI_WAVELENGTH}. We find that in the
innermost~$12{\rm M}\simeq 91(M_{\rm NS}/1.4 M_\odot)\,\rm km$ of the
disk and outside the ISCO, $\alpha_{\rm SS}$ ranges between~$0.01$
and~$0.03$ (see Table~\ref{table:summary_BHNSresults}), consistent
with values reported in earlier accretion disk
studies~\cite{Krolik2007,Gold:2013zma}. Therefore, it is expected that
magnetic turbulence driven by MRI is operating to some degree in our
simulations and drives the quasisteady accretion. However, farther
studies at higher resolution are required to confirm these results. We
compute the rest-mass accretion rate~$\dot{M}$ through
Eq.~A11~in~\cite{Farris:2009mt} and found after $\sim 1500M\simeq
38(M_{\rm NS}/1.4M_\odot)\,$ms following the merger, the accretion
begins to settle to a quasi-stationary state and decays slowly
afterward. In magnetized cases, we find that $\dot{M}$ is roughly
$0.3M_\odot/s$ (see Table~\ref{table:summary_BHNSresults}) once the
outflow reaches a height of $\sim 100M\simeq 760(M_{\rm
  NS}/1.4M_\odot)\,\rm km$.  At that time, the disk has a mass of
$\sim 0.13M_\odot(M_{\rm NS}/1.4M_\odot)$ for Mq3NSm0.17 and Mq3NS0.0,
and of $\sim 0.2M_\odot(M_{\rm NS}/1.4M_\odot)$ for Mq3NSp0.23 (see
Table~\ref{table:summary_BHNSresults}). Hence the disk (jet's fuel) is
expected to be accreted in $\Delta t\sim M_{\rm disk}/\dot{M}\sim
0.5-0.8(M_{\rm NS}/1.4M_\odot)\,\rm s$, consistent with the lifetime
of typical sGRBs~\cite{Berger2014}.

As pointed out in~\cite{prs15}, following tidal disruption, magnetic
winding and the MRI amplify the magnetic field and induce magnetic
pressure gradients above the BH poles that eventually overcome the
baryon ram pressure of the fall-back debris and drive an outflow
collimated by the magnetic field~(see bottom panels in
Fig.~\ref{fig:BHNS_case_q31_OmegaH} and
\ref{fig:BHNS_case_s0_sm05}). In Mq3NSm0.17 we find that an incipient
jet is launched after $\sim 3500M\simeq 88(M_{\rm NS}/1.4M_\odot)\,\rm
ms$ following merger, while in high prograde spin case Mq3NSp0.23 the
jet emerges after $\sim 5500M\simeq 138(M_{\rm NS}/1.4M_\odot)\, \rm
ms$. This delay time is not unexpected. As the NS spin increases, the
tidal disruption occurs farther out from the ISCO (see
Eq.~\ref{eq:tidal}) and so more material has larger specific angular
momentum and spreads out further. In the prograde NS spin case, the
effective ISCO is smaller than in the retrograde
case~\cite{Barausse:2009xi}. Thus, a larger fraction of this material
remains outside BH horizon and induces a more baryon-loaded
environment that survives for a longer time. As a fraction of this
material rains back, the matter density above the BH poles drops and the
magnetic-field pressure gradients are able to overcome this ram
pressure and finally launch a jet. It should be noted that the jet
launching may not be possible for all EOSs if the matter fall-back
timescale is longer than the disk accretion
timescale~\cite{Paschalidis:2016agf}.

We measure the level of the collimation of the jet through the funnel
opening angle $\theta_{\rm jet}$ defined as the polar angle at which
the Poynting flux drops to $50\%$ of its maximum~\cite{prs15}.  Based
on the angle distribution of the outgoing flux on the surface of a
coordinate sphere with radius $60M\sim 455(M_{\rm NS}/1.4M_\odot)\rm
km$ (see Fig.~13~in~\cite{Ruiz:2018wah}), we estimate that the opening
angle of the jet is $\sim 25^\circ-30^\circ$.

Following the emergence of the jet, we verify the outgoing material in
the funnel has specific energy $E=-u_0-1 > 0$ (asymptotic region)
and hence is unbound.  The characteristic maximum value of the Lorentz
factor reached in the outflow is~$\Gamma_L \sim 1.2-1.3$. However, as
pointed out in~\cite{Vlahakis2003}, fluid elements can be accelerated
to $\Gamma_L\simeq b^2/2\rho_0\simeq 100$ (see Table~
\ref{table:summary_BHNSresults}) consistent with sGRBs which
require~$\Gamma_L\gtrsim 20$~\cite{Zou2009}.

\begin{figure}
  \centering
  \includegraphics[width=0.49\textwidth]{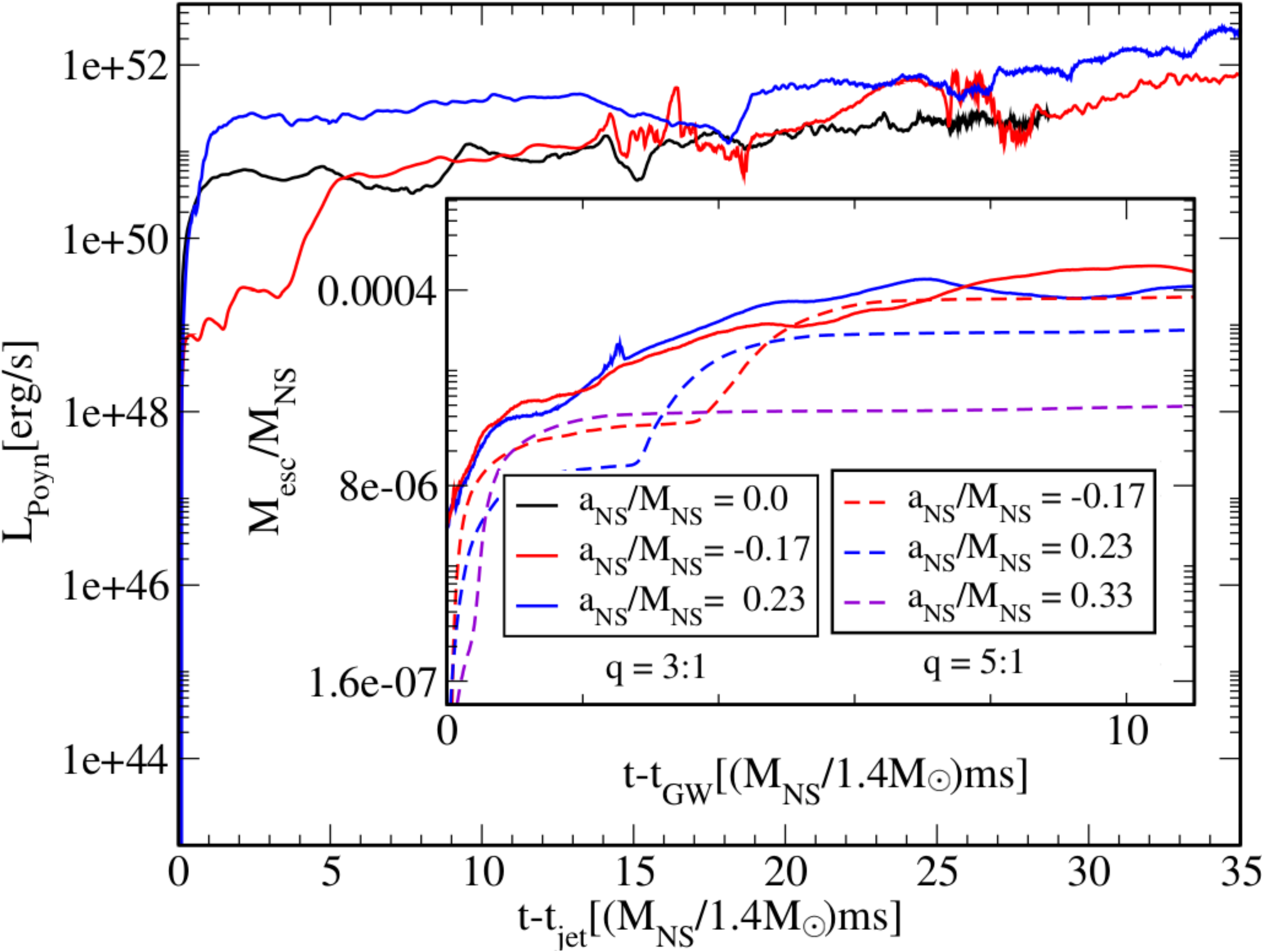}
  \caption{Outgoing EM (Poynting) luminosity following jet launching,
    computed on a coordinate sphere of radius $r=80{\rm M}\simeq
    606(M_{\rm NS}/1.4M_\odot)\,\rm km$ for the magnetized cases
    listed in Table~\ref{table:summary_BHNSresults}. The inset shows
    the rest-mass fraction of escaping matter following the peak
    amplitude of GWs.  
    \label{fig:Luminosity_ejecta}}
\end{figure}

Fig.~\ref{fig:Luminosity_ejecta} shows the outgoing EM (Poynting)
luminosity computed through $L_{\rm Poyn}\equiv -\int
T^{r(EM)}_t\,\sqrt{-g}\,d\mathcal{S}$~\cite{Ruiz:2019ezy} across a
spherical surface of coordinate radius $r=80{\rm M}\simeq 606(M_{\rm
  NS}/1.4M_\odot)\,\rm km$. The luminosity is $L_{Poyn}\simeq
10^{51.5\pm 0.5}\,\rm ergs/s$ (see
Table~\ref{table:summary_BHNSresults}), and hence is consistent with
the BZ luminosity~ $L_{\rm BZ}\sim 10^{51}\,a^2\,B^2_{15}\,
M_{5}^2\,\rm erg/s$ (see Eq. 4.50 in~\cite{Thorne86}), as well as with
typical sGRB (equivalent isotropic)
luminosities~(see~e.g.~\cite{2011ApJ...734...58S}).  Here
$B_{15}=B/10^{15}\,\rm G$ and $a=a_{\rm BH}/M_{\rm BH}$ and
$M_{5}=M_{\rm BH}/ 5M_\odot$.

To further assess if the BZ mechanism~is operating in the BHNS remnants we compute the
ratio of the angular velocity of the magnetic fields to the angular velocity of the BH
$\Omega_F$~(see Eq.12~ in~\cite{Ruiz:2018wah}) on a meridional plane passing through
the BH centroid and along semicircles of coordinate radii between~$r=R_{\rm BH}$
and~$2\,R_{\rm BH}$. In all cases~$\Omega_F$~ranges between~$\sim 0.4-0.45$ at the BH
poles and $\sim 0.1$ near the equator, and hence the field lines are differentially rotating.
It should be noted that deviations from the expected $\Omega_F=0.5$ value~(see e.g.
\cite{Komissarov2001}) can be attributed to the deviations from the gauge in which $\Omega_F$
is computed ($\Omega_F$ is defined for stationary and axisymmetric spacetimes in Killing
coordinates), deviation from a split-monopole magnetic field or lack of resolution~\cite{prs15}.

We measure the dynamical ejection of matter (ejecta) through $M_{\rm
  esc} = \int \rho_*\,d^3x$ outside a coordinate radius $r > r_0$, and
under the following conditions: a) $E=-1 - u_0 > 0$, and b) positive
(outgoing) radial velocity of the ejected material.  Here
$\rho_*\equiv-\sqrt{\gamma}\rho_0\,n_\mu\,u^\mu$, where $\gamma$ is
the determinant of the three metric and $n^\mu$ the future pointing
normal vector to a $t=$~constant hypersurface.  To verify that our
results are independent of $r_0$ at large radius, we compute the mass
of the ejecta varying $r_0$ between $30M\simeq 230~(M_{\rm
  NS}/1.4M_\odot)\,\rm km$ and $100M\simeq 760 (M_{\rm NS}/
1.4M_\odot)\,\rm km$. As shown in the inset of
Fig.~\ref{fig:Luminosity_ejecta}, the initial NS spin has a strong
effect on the ejecta (especially in the $q=5:1$ case that we discuss
in the next section). The ejecta in Mq3NSp0.23 is around $35\%$ higher
than in Mq3NSm0.17, where it turns out to be~$10^{-2.1}M_\odot(M_{\rm
  NS}/1.4M_\odot)$. Slightly smaller values of the ejecta have been
recently reported in BHNS mergers where the NS companion
(irrotational) is modeled with softer (H-type)
EOSs~\cite{Hayashi:2020zmn}. Ejecta masses $\gtrsim 10^{-3}M_\odot$
are expected to lead to detectable, transient kilonovae
signatures~(see e.g.~\cite{Metzger:2016pju}) powered by radioactive
decay of unstable elements formed by the neutron-rich material ejected
during BHNS mergers~\cite{Li:1998bw,
  Metzger:2016pju}. In~\cite{Barnes:2013wka} it was shown that the
opacities in r-process ejecta are likely dominated by lanthanides,
which induce peak bolometric luminosities for kilonovae
of~\cite{East:2015yea}
\begin{equation}
  L_{\rm knova}\approx 10^{41}\left(\frac{M_{\rm eje}}
  {10^{-2}M_{\odot}}\right)^{1/2}\,\left(\frac{v_{\rm eje}}{0.3c}\right)^{1/2}\,
  \rm erg/s\,,
 \label{L_peak_knove}
\end{equation} 
and rise times of~\cite{East:2015yea}
\begin{equation}
  t_{\rm peak}\approx0.25\,\left(\frac{M_{\rm eje}}{10^{-2} M_{\odot}}\right)^{1/2}\,
  \left(\frac{v_{\rm eje}}{0.3c}\right)^{-1/2}\, \rm days\,.
\label{t_peak_knove}
\end{equation}
Here $v_{\rm eje}$ and $M_{\rm eje}$ are the mass-averaged velocity
and rest-mass of the ejecta.  Using the above equations, we estimate
that the bolometric luminosity of potential kilonovae signals is
$L_{\rm knova}=10^{41.3\pm0.1}\rm erg/s$ with rise times of
$0.18-0.27$~days~(see~Table~\ref{table:summary_BHNSresults}).  These
luminosities correspond to an R band magnitude of $\sim 24$ mag at
$200$~Mpc (inside the aLIGO volume~\cite{Aasi:2013wya}), and above the
LSST survey sensitivity of $24.5$
mag~\cite{Barnes:2013wka,East:2015yea}, and hence may be detectable by
the LSST survey.

Finally, we compute the characteristic interior temperature $T_{\rm disk}$ of the
disk remnant assuming that the specific thermal energy density~$\epsilon_{\rm th}$
can be modeled as~\cite{Etienne:2008re}
\begin{equation}
  \epsilon_{\rm th} = \frac{3\,k_{\rm B}\,T_{\rm disk}}{2\,m_n} +f_{\rm s}\,
  \frac{a\,T_{\rm disk}^4}{\rho_0}\,,
\label{eq:temperature}
\end{equation}
where $k_{B}$ is the Boltzmann constant, $m_n$ is the mass of a
nucleon, and $a=8\pi^5 k_{\rm B}^4/(15\,h_{\rm P}^3)$ is the radiation
constant. Here $h_{\rm P}$ is the Plank constant. As pointed out
in~\cite{Etienne:2008re}, the first term in Eq.~\ref{eq:temperature}
is approximately the thermal energy of the nucleons, while the second
represents the thermal energy due to radiation and thermal
relativistic particles. The factor $f_{\rm s}$ accounts for the number
of species of ultrarelativistic particles that contribute to thermal
energy.  When $T \ll 10^{10}\,\rm K$ thermal radiation is dominated by
photons and $f_{\rm s}=1$. When $T\gg 10^{10}\,\rm K$, electrons and
positrons become ultrarelativistic and also contribute to radiation,
and hence $f_{\rm s}=1+2\times (7/8) = 11/4$.  At sufficiently high
temperatures ($T \gtrsim 10^{11}\,\rm K$) and
densities~($\rho_0\gtrsim 10^{12}\,\rm g/cm^3$), thermal neutrinos and
antineutrinos are copiously generated and become trapped. Taking into
account three flavors of neutrinos and antineutrinos $f_{\rm s}=11/4 +
3\times (7/8) = 43/8$.

We measure the thermal energy generated by shocks through the ratio
$K=P/P_{\rm cold}$ (entropy parameter), where $P_{\rm cold}=\kappa
\rho_0^\Gamma$ is the pressure associated with the cold EOS used to
build our initial configurations. In all cases we find the
characteristic value of the entropy parameter in the disk is $K\sim
200$. Next, we compute the specific thermal energy as~$\epsilon_{\rm
  th} =(K-1)\,\epsilon_{\rm cold}$ with $\epsilon_{\rm cold}=
\kappa\,\rho_0$ for a polytropic EOS with $\Gamma=2$ (see
Eq.~12~in~\cite{Etienne:2008re}).  Plugging these values in
Eq.~\ref{eq:temperature}, we find the characteristic value of the
temperature in the disk is $T_{\rm disk}\sim 10^{11.0}\,\rm K$
(or~8.6~MeV) for the nonmagnetized cases, where the characteristic
densities in the disk are~$\rho_0\sim 10^{12}\,\rm gm/cm^3$ (see
bottom panels in Fig.~\ref{fig:BHNS_q31:hydro}), and $T\sim
10^{10.6}\,\rm K$ (or~3.4~MeV) for the magnetized cases, where the
characteristic densities are $\rho_0\sim 10^{11}\,\rm gm/cm^3$ (see
bottom panels in Figs.~\ref{fig:BHNS_case_q31_OmegaH}
and~\ref{fig:BHNS_case_s0_sm05}).  Thus, these hot accretion disks may
emit a copious amount of neutrinos with a peak luminosity
of~$10^{53}\,\rm erg/s$ through thermal pair production and subsequent
electron/positron captures on free
nucleons~\cite{Kyutoku:2017voj}. However, their lifetimes might be too
small to explain the majority of sGRBs~\cite{Just:2015dba}. It has
been suggested that BH + disk engines that power typical sGRBs may be
dominated initially by thermal pair production followed by the BZ
process, leading to a transition from a thermally dominated fireball
to a Poynting dominated outflow as observed in some GRBs, such as GRB
160625B~\cite{Dirirsa:2017pgm}.

\begin{figure}
  \centering
  \includegraphics[width=0.48\textwidth]{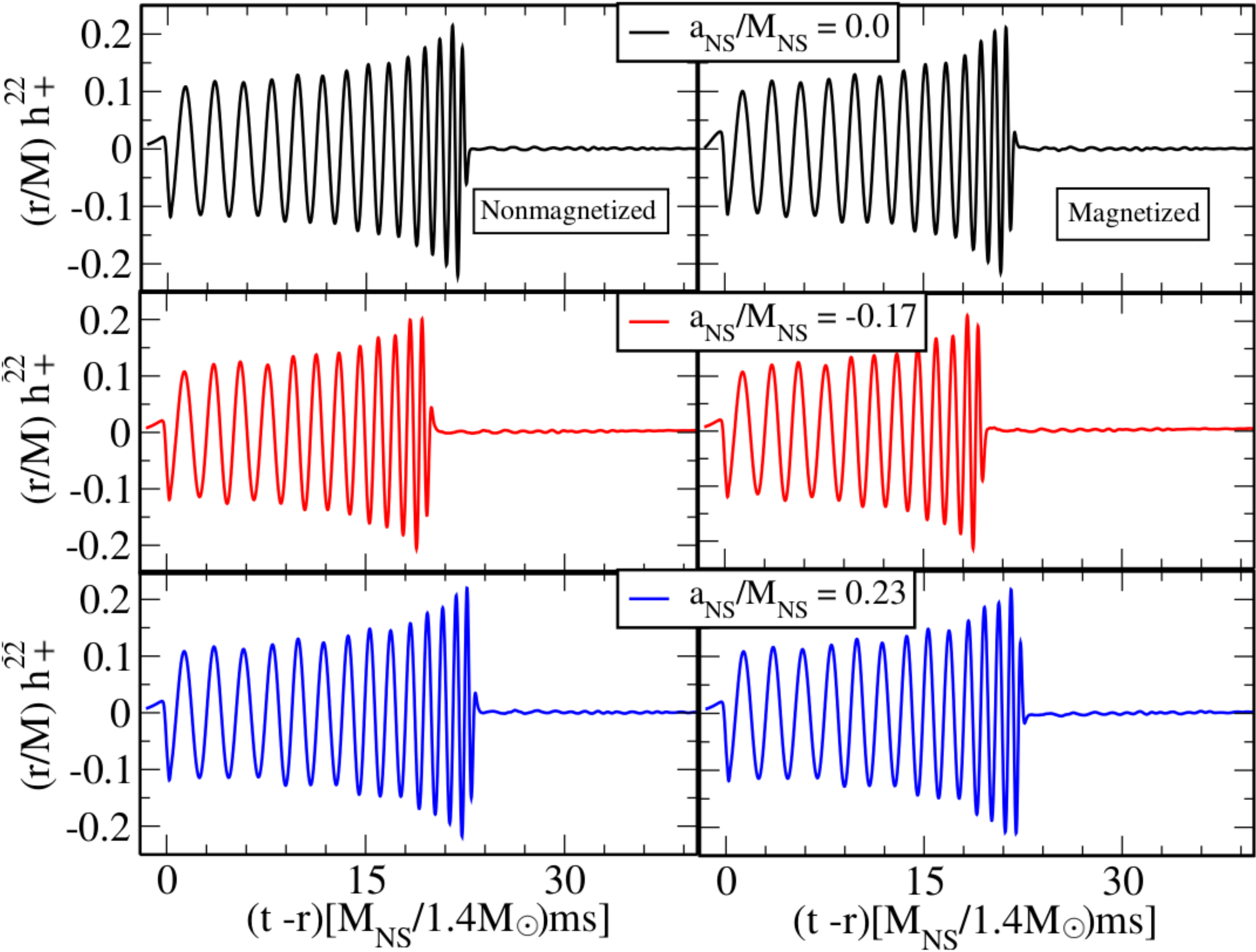}
  \caption{Mode $(l,m)=(2,2)$ of the GW strain $h_+$ as function of the retarded time
    extracted at a coordinate sphere of radius $r=80M\sim 606(M_{\rm NS}/1.4M_\odot)\,$km
    for nonmagnetized (left column) and magnetized (right column) cases with mass ratio
    $q=3:1$~(see~Table~\ref{table:summary_BHNSresults}).
    \label{fig:GWs_q31}}
\end{figure}

Figure~\ref{fig:GWs_q31} shows the GW strain $h_+$ of the dominant
mode $(2,2)$ for these configurations. Left column displays the
nonmagnetized evolutions, while the right one displays the magnetized
evolutions. The corresponding binaries (nonmagnetized and magnetized)
merge roughly at the same time (here the merger time $t_{\rm mer}$~is
defined as the time of peak amplitude of the GWs). This result is
anticipated because the seed magnetic field is dynamically weak and
there is no significant enhancement of its magnitude during the
inspiral~(the seed magnetic field is simply advected with the
fluid). However, due to the hang-up effect~\cite{clz06}, the prograde
NS spin configuration~($a_{\rm NS}/M_{\rm NS}=0.23$) aligned with the
total orbital angular momentum of the system (bottom~panel in~Fig.
\ref{fig:GWs_q31})~undergoes about one or two more orbits compared to
the irrotational and the retrograde NS spin ($a_{\rm NS}/M_{\rm
  NS}=-0.17$) cases (top and middle panels), respectively. A similar
effect has been reported in BBHs~\cite{clz06}, and
BNSs~\cite{Tsatsin2013,Kastaun2013,Dietrich:2016lyp,Ruiz:2019ezy,Tsokaros:2019anx,East:2019lbk}.

%
%

%

\subsection{Mass ratio $q=5:1$}
\label{sec:q=5_sec}
%
\begin{figure}
  \centering
  \includegraphics[width=0.49\textwidth]{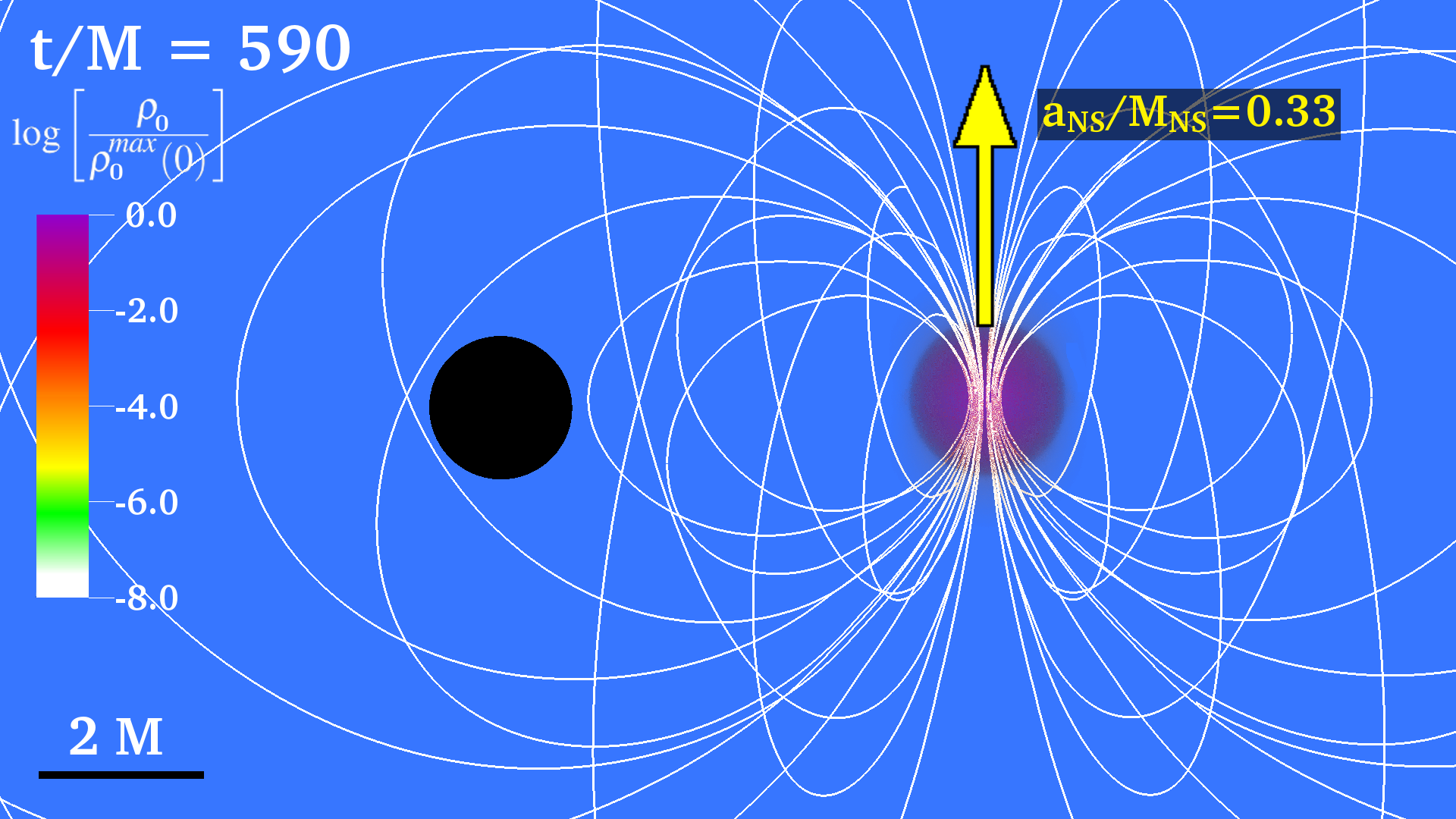}
  \includegraphics[width=0.49\textwidth]{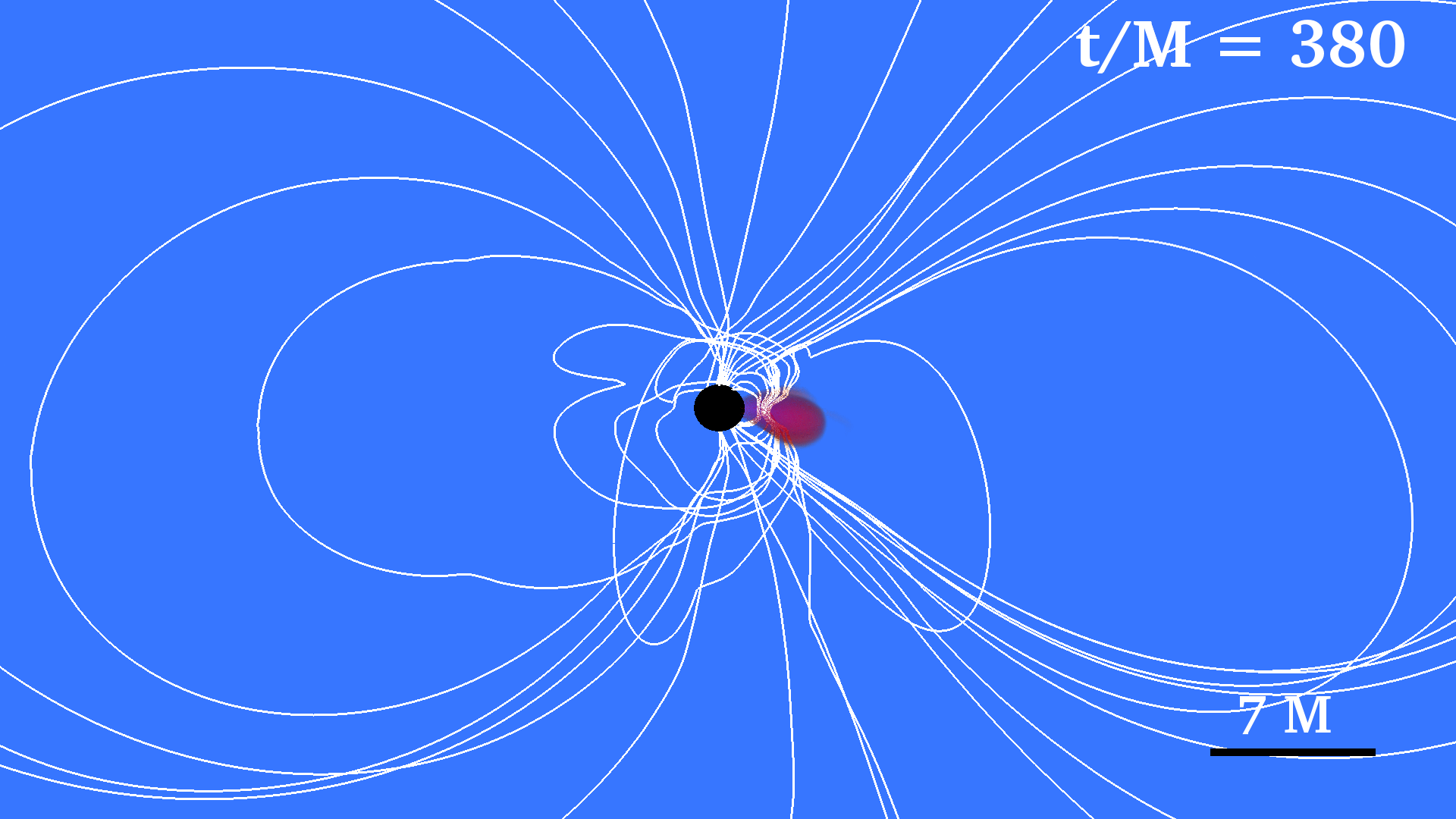}
  \includegraphics[width=0.49\textwidth]{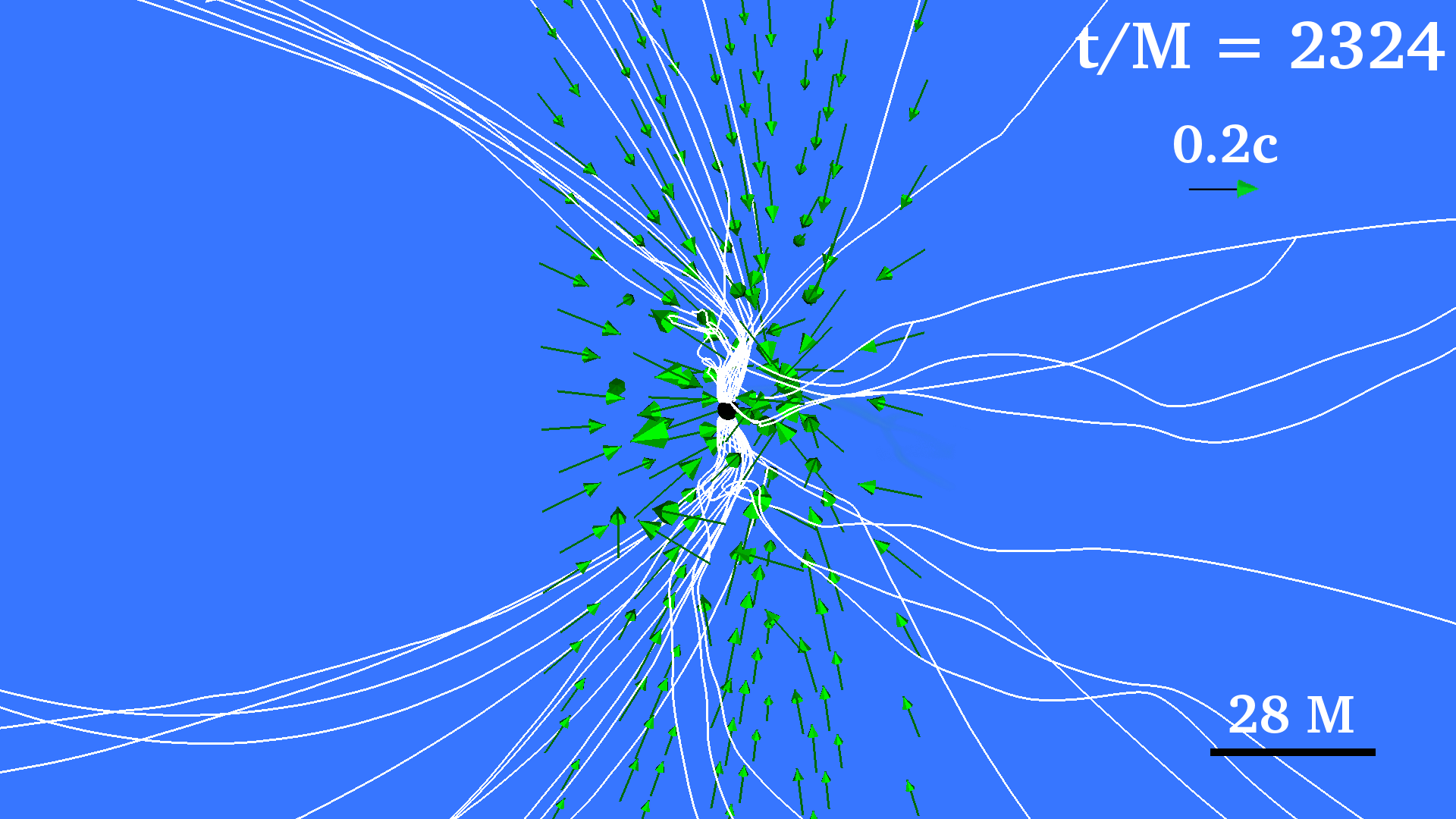}
  \caption{Same as Fig.~\ref{fig:BHNS_case_q31_OmegaH} but for 
    Mq5NSp0.33.
    \label{fig:q=5:1}}
\end{figure}
Fig.~\ref{fig:q=5:1} summarizes the evolution of Mq5NSp0.33 (our
extreme case). All configurations with mass ratio $q=5:1$ in
Table~\ref{table:summary_BHNSresults} have basically the same fate
independent of the magnitude of the magnetic field or the initial NS
spin: The tidal disruption occurs closer to the ISCO, resulting in
short tidal tails (see middle panel)~that leave stellar debris outside
the BH horizon with mass $\lesssim 1.4\%$ of the rest-mass of the NS
(bottom panel of~Fig.~\ref{fig:M0_outside}). See
Table~\ref{table:summary_BHNSresults} for other cases.
%
\begin{figure}
  \centering
  \includegraphics[width=0.50\textwidth]{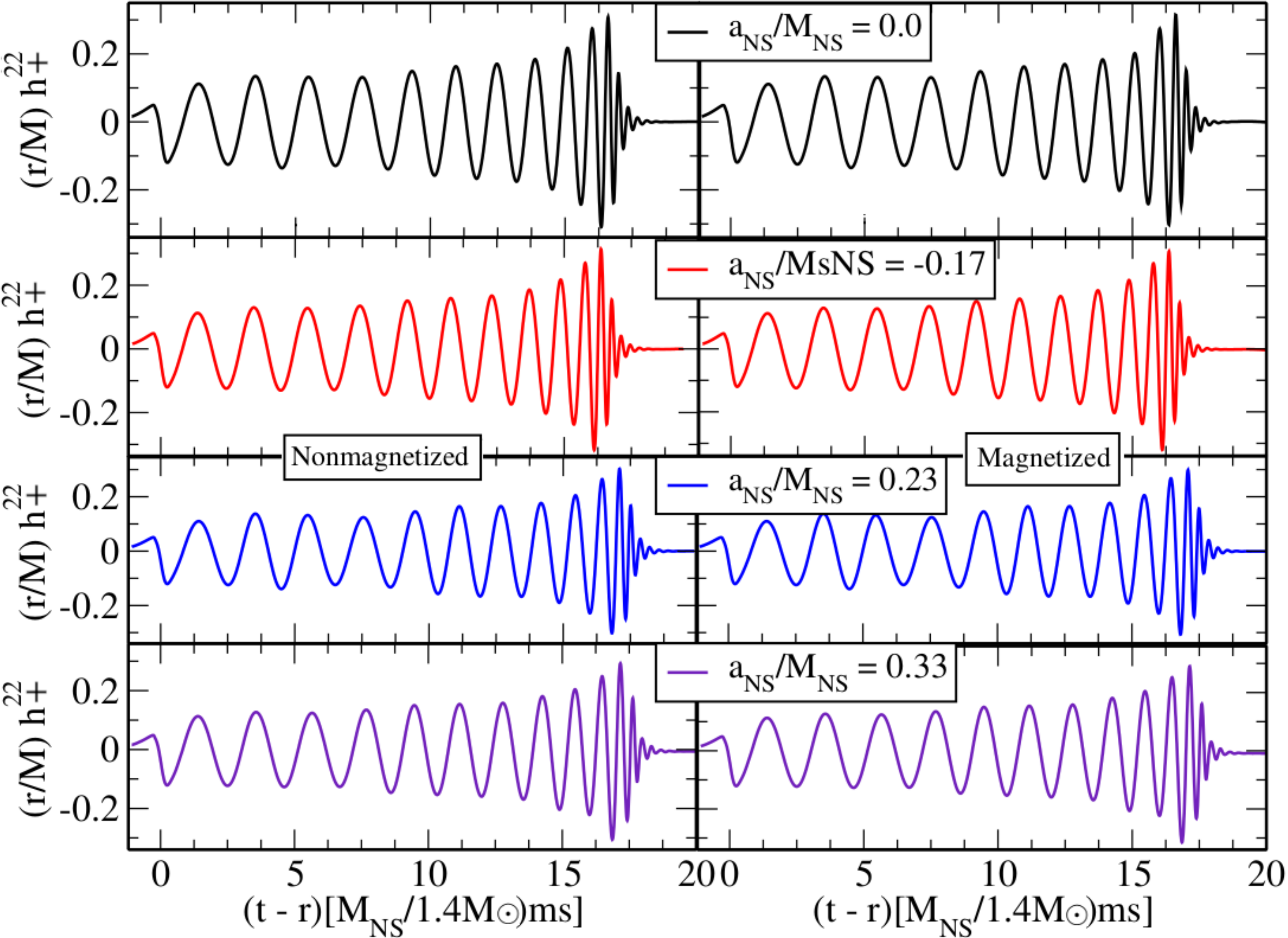}
  \caption{Same as Fig.~\ref{fig:GWs_q31} but for BHNS configuration with mass ratio~$q=5:1$
    (see~Table~\ref{table:summary_BHNSresults}).
    \label{fig:GWs_q51}}
\end{figure}
%
We do not find evidence for an outflow or large-scale magnetic field collimation (see bottom
panel in~Fig.~\ref{fig:q=5:1}).

Bottom panel of~Fig.~\ref{fig:M0_outside} shows the rest-mass fraction
outside the BH horizon.  We observe that in Mq5NSp0.33 the mass
outside the BH horizon is around three times larger than in
Mq5NSm0.17. So, the higher the prograde NS spin, the larger the
fraction of matter ejected. The inset of
Fig.~\ref{fig:Luminosity_ejecta} shows the dynamical ejection of
matter for the magnetized cases (similar values are found for the
nonmagnetized cases). In the extreme cases, the ejecta ranges between
$10^{-4.5}M_\odot(M_{\rm NS}/1.4M_\odot)$ (for Mq5NSm0.17)~and
$10^{-3.5}M_\odot( M_{\rm NS}/1.4M_\odot)$~(for Mq5NSp0.33). The
latter is near the threshold value of ejecta required to give rise to
a detectable kilonova~\cite{Metzger:2016pju}. Using
Eqs.~\ref{L_peak_knove} and~\ref{t_peak_knove} we estimate peak
bolometric luminosities from potential kilonovae of $L_{\rm knova}\sim
10^{40.0}-10^{40.6}$ erg/s and rise times $\lesssim 1~$h~(see~Table
\ref{table:summary_BHNSresults}). Such a kilonova is potentially
detectable by the LSST survey, although it would require rapid
response and high cadence EM follow-up observations. These results
suggest that even in the absence of a jet, the GWs from BHNS mergers
with moderate mass ratio and/or moderate BH spin may be accompanied by
detectable kilonovae signatures if the companion is a highly spinning
NS, which agrees with~\cite{East:2015yea}.

The GW strain $h_+$ of the dominant mode $(l,m)=(2,2)$ for the
nonmagnetized (left column) and magnetized (right column) evolutions
is shown in~Fig.~\ref{fig:GWs_q51}. We observe that the more extreme
configurations ($a_{\rm NS}/M_{\rm NS}=0.23$ and $a_{\rm NS}/M_{\rm
  NS}=0.33$) displayed on the two bottom rows undergo about half an
orbit ($a_{\rm NS}/M_{\rm NS}=0.23$) and a full orbit ($a_{\rm
  NS}/M_{\rm NS}=0.33$) more compared to the nonspinning and
retrograde NS spin cases (first two rows). Thus, NS spin can lead to
dephasing and should be accounted for in BHNS waveform templates.

\subsection{Distinguishability of the gravitational waves}
\label{subsec:GW_and_LIGO}
%
\begin{figure}
  \centering
  \includegraphics[width=0.50\textwidth]{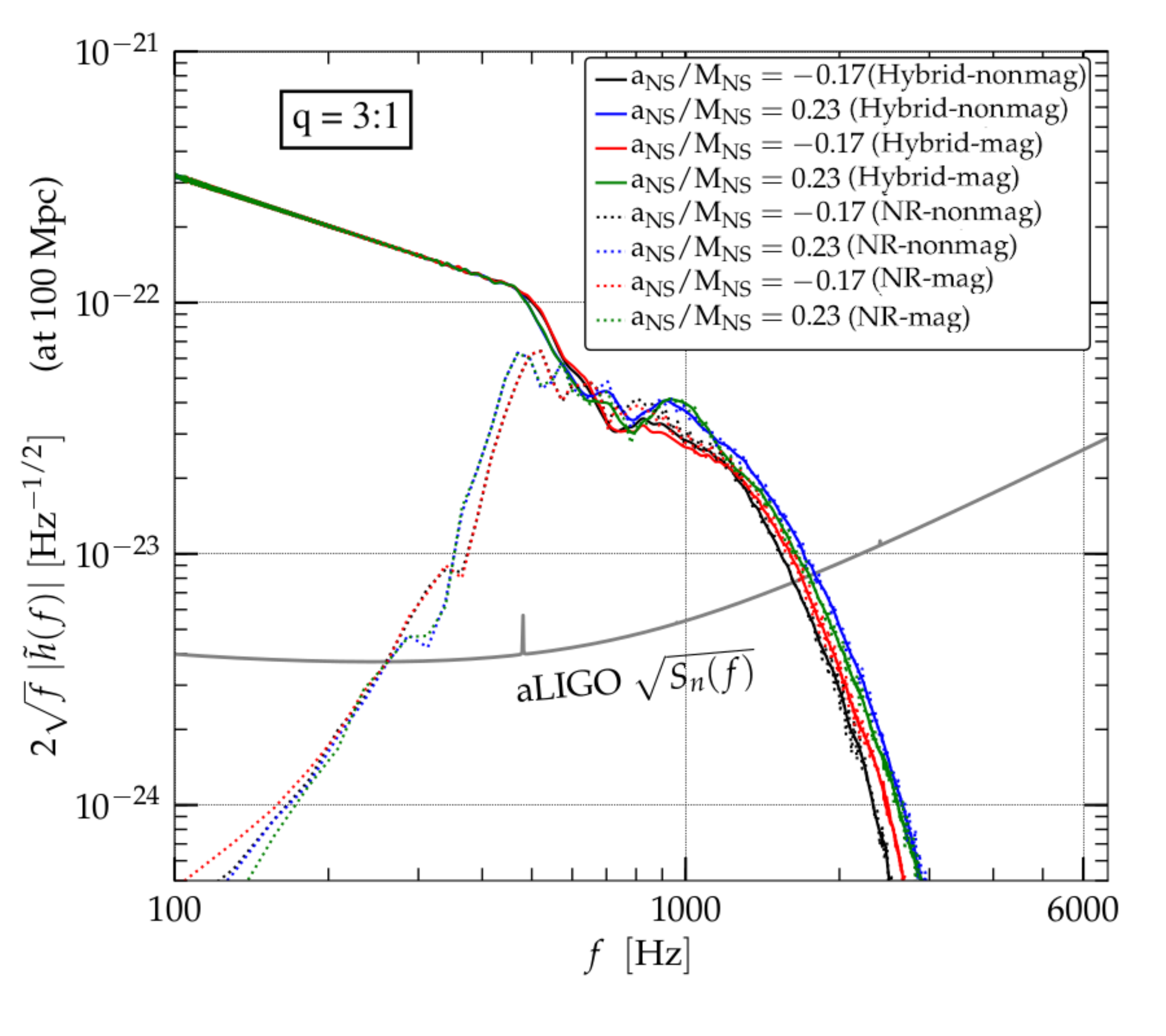}
  \includegraphics[width=0.50\textwidth]{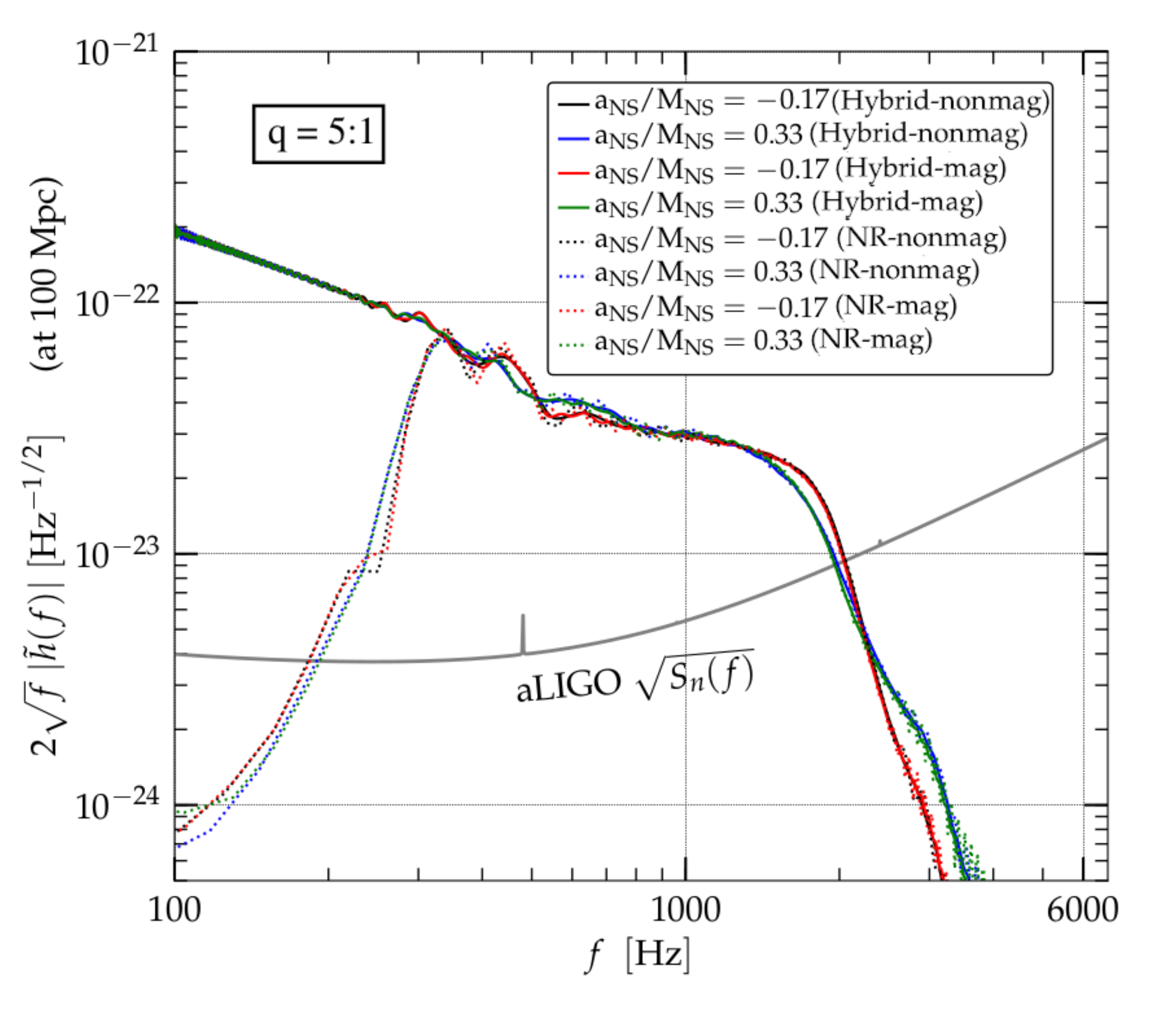}
  \caption{Gravitational-wave power spectrum of the dominant mode $(l,m)=(2,2)$
    at a source distance of $100\rm Mpc$ for our extreme cases with mass ratio $q=3:1$
    (top panel) and mass ratio $q=5:1$ (bottom panel),
    along with the aLIGO noise curve. This curve corresponds to the {\tt ZERO$\_$DET$\_$HIGH$\_$P}
    configuration~\cite{Shoemaker09L}.
    Solid curves display the hybrid waveform found by appending the TaylorT1 PN
    waveform to the raw numerical signal (dotted curves).
    \label{fig:fft}}
\end{figure}
As mentioned in the previous section, as the initial NS spin
increases, the binary inspiral lasts longer, resulting in more
gravitational wave cycles (see Figs.~\ref{fig:GWs_q31} and
\ref{fig:GWs_q51}). 
This enhancement induces a dephasing with respect to the non-spinning
cases, and a change in the amplitude of the GWs of $\lesssim 5\%$
between the respective waveforms, which is reflected in slight changes
in the energy $\Delta E_{\rm GW}$ and angular momentum $\Delta J_{\rm
  GW}$ carried away by GWs (see
Table~\ref{table:summary_BHNSresults}), as well as in the kick
velocity $v_{\rm kick}$ due to recoil~(see~Eqs.~3.7 and
3.20~in~\cite{Ruiz:2007yx}).  In this section, we probe if these
spin-driven and magnetic effects can be distinguished by aLIGO.

We start by extending the GW spectra in the frequency domain, creating
a hybrid waveform.  Following~\cite{Etienne:2011ea}, we append a
TaylorT1 post-Newtonian waveform~\cite{nijidataformat} to that of our
numerical relativity simulations. The hybrid waveform is then obtained
by minimizing
\begin{equation}
  \bigintsss_{t_{\rm i}}^{t_{\rm f}} dt\,
         \left[(h_+^{\rm NR} -h_+^{\rm PN})^2 +
          (h_x^{\rm NR} -h_x^{\rm PN})^2\right]^{1/2}\,,
\label{eq:hybrid}
\end{equation}
via the Nelder-Mead algorithm~\cite{Nelder:1965zz}, using as free
parameters the initial PN phase, amplitude, and orbital angular
frequency. In cases with mass ratio $q=3:1$ we integrate the above
expression between $t_{\rm i}= 150M$~and $t_{\rm f}=450M$, while in
cases with mass ratio $q=5:1$ the integration range is between $t_{\rm
  i}= 150M$ and~$t_{\rm f}=250M$.

Fig.~\ref{fig:fft} shows the GW spectrum of the dominant mode
$(l,m)=(2,2)$ at a source distance of $100\rm Mpc$ for our extreme
cases listed in Table~\ref{table:summary_BHNSresults}, along with the
aLIGO noise curve of the {\tt ZERO$\_$DET$\_$HIGH$\_$P}
configuration~\cite{Shoemaker09L}.  Solid (dotted) lines display the
hybrid (raw numerical) signals. We observe that the GW spectrum for
cases with mass ratio $q=3:1$ (top panel) rapidly decays as the star
becomes disrupted ($f\sim 900~(M_{\rm NS}/1.4M_\odot)^{-1}\,\rm
Hz$). By contrast, in cases with mass ratio $q=5:1$ (bottom panel),
the signal slowly decays until merger, where the star basically
plunges into the BH ($f\sim 1700~(M_{\rm NS}/1.4 M_\odot)^{-1}\,\rm
Hz$), and then the GW power drops significantly. Unlike the NS spin
imprints on the GWs evident at high frequencies, the magnetic field
imprints are not significant. In the lowest mass ratio cases, the
differences in the GW spectrum are marginally observable in the aLIGO
band~(top panel), while on those with mass ratio $q=5:1$ are not
evident at all even outside the aLIGO band~(bottom panel).

A more precise way to assess the distinguishability is through the match function
$\mathcal{M}_{\text{\tiny{GW}}}$ defined as~\cite{Allen2012}
\begin{equation}
\mathcal{M}_{\text{\tiny{GW}}} =  \underset{(\phi_c,t_c)}{{\rm max}}
\frac{\left<{h}_1|{h}_2(\phi_c,t_c)\right>}{\sqrt{\left<{h}_1|{h}_1\right>
    \left<{h}_2|{h}_2\right>}}\,,
\label{eq:match}
\end{equation}
between two given waveforms. The maximization is taken over a large set
of phase shifts $\phi_c$ and time shifts $t_c$. Here $\left<{h}_1|{h}_2\right>$ denotes
the  noise-weighted inner product~\cite{Allen2012}
\begin{equation}
\left<h_1|h_2\right>= 4\,{\rm Re}\int_0^\infty\frac{\tilde{h}_1(f)\,\tilde{h}^*_2(f)}
  {S_h(f)}\,df\,,
\end{equation}
were $h=h_+-i\,h_\times$, $\tilde{h}$ is the Fourier transform of the
strain amplitude $\sqrt{\tilde h_+(f)^2 +\tilde h_{\times}(f)^2}$ of
the dominant mode $(l,m)=(2,2)$, and $S_h(f)$ is the power spectral
density of the aLIGO noise~\cite{Shoemaker09L}. In our extreme cases,
we find that $\mathcal{M}_{\text{\tiny{GW}}}=0.9980$ between the
waveforms of the q3NSm0.17 cases, and
$\mathcal{M}_{\text{\tiny{GW}}}=0.9982$ between those of the
q3NSp0.23.  Similarly, $\mathcal{M}_{\text{\tiny{GW}}}=0.9998$ for all
of our extreme cases with mass ratio~$q=5:1$.

The standard choice for the threshold match for distinguishing two
signals is $1-1\slash (2\rho^{2})$, where $\rho$ here is the
signal-to-noise ratio (SNR). For a SNR of 15~\cite{Harry_2018}, two
signals are distinguishable when
$\mathcal{M}_{\text{\tiny{GW}}}\lesssim 0.9956$. Using Equation~(18)
in~\cite{Baird2013} (with one degree of freedom since we compare
configurations varying one parameter), a match of $1-1\slash
(2\rho^{2})$ corresponds to a 68\% confidence level. Thus, it seems
unlikely that aLIGO detectors can detect the magnetic field imprints
on the GWs even at 68\% confidence level. By contrast, the imprints of
the NS spin on the waveform are more easily detectable. The match
function between the waveform of Mq3NSp0.23 and Mq3NSm0.17
is~$0.9844$, and $0.9874$~between those of Mq5NSp0.33 and Mq5NSm0.17,
and hence smaller than the threshold match value.

%
\section{Conclusions}
\label{sec:conclusion}
We previously reported self-consistent MHD numerical simulations in
full GR showing that BHNS binaries undergoing merger and significant
NS tidal disruption outside the ISCO can launch a magnetically-driven
jet~\cite{prs15,Ruiz:2018wah}. This happens whenever a net poloidal
magnetic flux is accreted onto the BH with $B^2/8\pi\rho_0 \gg 1$
above the BH poles, and hence these systems serve as a possible
progenitor of the central engine that powers a sGRB. However,
population synthesis studies \cite{Belczynski:2007xg,bdbofh10} along
with the reported aLIGO/Virgo GW
detections~\cite{TheLIGOScientific:2017qsa,alertsGWweb,Abbott:2020niy}
suggest that in typical BHNSs we may have $M_{\rm BH}/M_{\rm NS}
\gtrsim 5$. For such high mass ratios the NS companion simply plunges
into the BH before undergoing tidal disruption, leaving a negligible
amount of matter outside the BH horizon~($\lesssim 2\%$ of the NS
rest-mass). Thus, BHNS mergers may not be accompanied by a near
simultaneous, observable EM counterpart. However, the NS spin can have
a strong impact on the dynamical ejection of matter, and so may lead
to subsequent kilonovae signatures. It should be noted that the
current GW observations set constraints on the effective spin of the
binary but not on the individual spins.

In this paper, we explored the impact of the NS spin companion on the
dynamical ejection of matter, the mass of the accretion disk, and the
jet launching, from BHNS binaries undergoing merger with moderate mass
ratios ($q=3:1$ and $q=5:1$). For comparative purposes, we considered
the quasiequilibrium BHNS initial data used previously
in~\cite{Ruiz:2018wah,Etienne:2008re} and endowed the NS companion
with an effective spin~\cite{Ruiz:2014zta}.

Consistent with our previous results~\cite{prs15,Ruiz:2018wah}, we
found that all magnetized cases with mass ratio $q=3:1$~in
Table~\ref{table:summary_BHNSresults} launch a magnetically-driven jet
after $\Delta t\sim 3500M-5500\approx 88-138(M_{\rm NS}/
1.4M_\odot)\,\rm ms$ following the peak GW signal (see bottom panels
in Figs.  \ref{fig:BHNS_case_q31_OmegaH}
and~\ref{fig:BHNS_case_s0_sm05}). At these times the force-free
parameter above the poles of the BH reaches values of
$b^2/(2\,\rho_0)\gtrsim 100$ (~see
Table~\ref{table:summary_BHNSresults}). The time delay between GW peak
and jet launching depends strongly on the NS spin. The larger the NS
spin, the longer the delay. This result can be explained by the fact
that as the prograde NS spin increases the effective
ISCO~\cite{Barausse:2009xi} decreases, while the NS becomes less
bound, so that the onset of NS tidal disruption occurs farther out
from the ISCO. This causes long tidal tails of matter having larger
specific angular momentum that spread out and form a baryon-loaded
environment that persists for a longer time. The lifetime of the jet
[$\Delta t\sim 0.5-0.8 (M_{\rm NS}/1.4M_\odot)\rm s$] and outgoing
Poynting luminosity~[$L_{\rm Poyn}\sim 10^{51.5\pm 0.5}\,\rm erg/s$]
are consistent with typical sGRBs, as well as with the BZ
mechanism~\cite{BZeffect}.
Consistent with our previous results~\cite{prs15,Ruiz:2018wah}, we
estimated that the opening angle of the jet is $\sim
25^\circ-30^\circ$. In contrast to the $q=3:1$ cases, we do not find
evidence of outflow or large-scale magnetic field collimation in
any of the BHNS cases with mass ratio~$q=5:1$, (see bottom panel in
Fig.~\ref{fig:q=5:1}). Persistent fall-back debris in the atmosphere
are observed until the termination of our simulations.

We estimated the characteristic interior temperature of the disk
remnant via Eq.~\ref{eq:temperature}, and found that it is $T_{\rm
  disk}\sim 10^{11}\,\rm K$ (or~8.6~MeV). Thus, it may emit a copious
amount of neutrinos with peak luminosity of~$10^{53}\,\rm
erg/s$~\cite{Kyutoku:2017voj}. It has been suggested that, as neutrino
annihilation may carry away a significant amount of energy from inner
regions of the disk, it may help the jet development. Thus, a BH +
disk remnant that powers a typical sGRB may be dominated initially by
thermal pair production followed by the BZ
process~\cite{Dirirsa:2017pgm}.

We observed that the dynamical ejection of matter is strongly affected
by the initial NS spin.  The ejecta ranges between $\sim
10^{-4.5}(M_{\rm NS}/1.4M_\odot)M_\odot$ and $\sim 10^{-2} (M_{\rm
  NS}/1.4M_\odot)M_\odot$, and may induce kilonovae signatures with
peak bolometric luminosities~of~$L_{\rm knova}\sim
10^{40}-10^{41.4}$~erg/s and rise times $\lesssim 6.5\,\rm h$,
potentially detectable by the LSST survey~\cite{Mandelbaum:2018ouv}.
These preliminary results suggest that moderately high-mass ratio BHNS
binaries undergoing merger, where the NS companion has a significant
spin, may give rise to a detectable kilonovae signatures even if a
magnetically-driven jet is absent.

Furthermore, we probed if magnetic-field and NS spin imprints on the
GWs can be distinguished by aLIGO. To assess this possibility, we
stitched a TaylorT1 Post-Newtonian waveform to that of our numerical
simulations. Next, we computed the GW power spectrum and the match
function~$\mathcal{M}_{\text{\tiny{GW}}}$~(see
Eq.~\ref{eq:match}). Unlike the NS spin imprints on the GWs evident at
high frequencies (see Fig.~\ref{fig:fft}), we found that in the lowest
mass ratio cases, the GW power spectrum of the corresponding
nonmagnetized and magnetized signals differ marginally inside the
aLIGO band, while for those with mass ratio $q=5:1$, the differences
are negligible even outside the aLIGO band. The match function for our
extreme cases is $\mathcal{M}_{\text{\tiny{GW}}}=0.9980$ between the
waveforms of q3NSm0.17 cases, and
$\mathcal{M}_{\text{\tiny{GW}}}=0.9982$ between those of
q3NSp0.23. Similarly, $\mathcal{M}_{\text{\tiny{GW}}}=0.9998$ for all
of our extreme case with mass ratio $q=5:1$.  It appears unlikely that
aLIGO detectors can detect magnetic-field imprints on the GWs which
require $\mathcal{M}_{\text{\tiny{GW}}}\lesssim 0.9956$ for a
signal-to-noise ratio of 15~\cite{Harry_2018}. By contrast, the
imprints of the NS spin on the waveform are more easily
detectable. The match function between the waveform of Mq3NSp0.23 and
Mq3NSm0.17 is~$0.9844$, and $0.9874$~between those of Mq5NSp0.33 and
Mq5NSm0.17.

Notice that due to the finite computational resources at our disposal, we
explored only two extreme cases to probe the impact of the NS spin on
the dynamical ejection of matter when the NS is (or is not) tidally
disrupted before merger. Our motivation for considering a
non-spinning BH in the $q=5:1$ case was to observe if the NS spin alone
can change the amount of mass left outside the BH horizon. Using the fitting
model in~\cite{Foucart:2012nc} (which applies to non-spinning BHs) for a
BHNSs with mass ratio $q=5:1$, a BH spin larger than $0.6$ is required
to tidally disrupt the NS before merger and to leave matter
outside the BH. Thus, by considering a worst case scenario for
ejecta and matter outside the BH we can study the NS spin effects
alone. We will explore more general cases in the future.

Finally, some caveats are in order. Despite the fact that our initial
data with NS spin slightly violate the constraints, as the evolution
proceeds our efficient constraint damping scheme decreases the
constraint violations down to the same levels as our
constraint-satisfying initial data with irrotational neutron stars.
However, some of the reported results in this work may be affected by
the fact that our method for endowing the NS with spin throws the NS
slightly off equilibrium for the highest spin values we
consider. While the trends with NS spin we reported are robust against
this property of our initial data, given that they are supported by
analytic arguments, the precise values of ejecta and disk masses we
reported could be affected. We plan to address these points in future
work.


\acknowledgements We thank the Illinois Relativity group REU team
(K. Nelli, M. N.T Nguyen, and S. Qunell) for assistance with some of
the visualizations. This work was supported by NSF Grants
No. PHY-1662211 and No. PHY-2006066, and NASA Grant No. 80NSSC17K0070
to the University of Illinois at Urbana-Champaign, and NSF Grant
PHY-1912619 to the University of Arizona. This work made use of the
Extreme Science and Engineering Discovery Environment (XSEDE), which
is supported by National Science Foundation Grant
No. TG-MCA99S008. This research is also part of the Frontera computing
project at the Texas Advanced Computing Center. Frontera is made
possible by National Science Foundation award OAC-1818253. Resources
supporting this work were also provided by the NASA High-End Computing
(HEC) Program through the NASA Advanced Supercomputing (NAS) Division
at Ames Research Center.
%
\bibliographystyle{apsrev4-1}        
\bibliography{references}            
\end{document}